\documentclass[11pt,a4paper]{article}
\pdfoutput=1
\usepackage{jcappub}
\allowdisplaybreaks
\usepackage{graphicx,psfrag}% Include figure files
\usepackage{bm}% bold math
\usepackage{mathbbol,verbatim}
\usepackage{slashed}
\usepackage{graphics}

\usepackage{tikz}
\usetikzlibrary{trees}
\usetikzlibrary{decorations.pathmorphing}
\usetikzlibrary{decorations.markings}
\usetikzlibrary{decorations, decorations.markings, decorations.pathmorphing, arrows, graphs, shapes.geometric, snakes}
\usetikzlibrary{arrows}

\tikzset{
photon/.style={decorate, decoration={snake}, draw=red},
dark/.style={draw=gray, postaction={decorate},
        decoration={markings,mark=at position .55 with {\arrow[draw=gray]{>}}}},
antidark/.style={draw=gray, postaction={decorate},
        decoration={markings,mark=at position .55 with {\arrow[draw=gray]{<}}}},
electron/.style={draw=violet, postaction={decorate},
        decoration={markings,mark=at position .55 with {\arrow[draw=violet]{>}}}},
neutrino/.style={draw,color=violet,thick, postaction={decorate} },
neutrinolight/.style={draw=blue, postaction={decorate} },
quark/.style={draw=blue, postaction={decorate},
        decoration={markings,mark=at position .55 with {\arrow[draw=blue]{>}}}},
antiquark/.style={draw=blue, postaction={decorate},
        decoration={markings,mark=at position .55 with {\arrow[draw=blue]{<}}}},
heavyquark/.style={draw=purple, postaction={decorate},
        decoration={markings,mark=at position .55 with {\arrow[draw=purple]{>}}}},
antiheavyquark/.style={draw=purple, postaction={decorate},
        decoration={markings,mark=at position .55 with {\arrow[draw=purple]{<}}}},
        gluon/.style={decorate, draw=or,
        decoration={coil,amplitude=2pt, segment length=3pt}},
gluon/.style={decorate, draw=or,
        decoration={coil,amplitude=2pt, segment length=3pt}},
ZZ/.style={decorate, decoration={snake,amplitude=1.5pt, segment length=5pt}, draw=greeen},
left,
  }

\definecolor{greeen}{rgb}{0.03,0.84,0.13}
\definecolor{test}{rgb}{0.03,0.74,0.33}
\definecolor{viol}{rgb}{0.44,0,0.94}
\definecolor{or}{rgb}{0.95,0.65,0}

\begin{document}

\begin{flushright}
ULB-TH/16-10
\end{flushright}

\title{Heavy right-handed neutrino dark matter and PeV neutrinos at IceCube}

\author[a]{P. S. Bhupal Dev,}
\author[b]{D. Kazanas,}
\author[c]{R. N. Mohapatra,}
\author[b,d]{V. L.  Teplitz,}
\author[e,f]{Yongchao Zhang}
\affiliation[a]{ Max-Planck-Institut f\"{u}r Kernphysik, Saupfercheckweg 1, D-69117 Heidelberg, Germany}
\affiliation[b]{Astrophysics Science Division, NASA Goddard Space Flight
Center, Greenbelt, MD 20771, USA}
\affiliation[c]{Maryland Center for Fundamental Physics, Department of Physics, University of Maryland, College Park, MD 20742, USA}
\affiliation[d]{Department of Physics, Southern Methodist University, Dallas, TX 75205, USA}
\affiliation[e]{Service de Physique Th\'{e}orique, Universit\'{e} Libre de Bruxelles, Boulevard du Triomphe, CP225, 1050 Brussels, Belgium}
\affiliation[f]{School of Physics, Sun Yat-Sen University, Guangzhou 510275, China}

\emailAdd{bhupal.dev@mpi-hd.mpg.de}
\emailAdd{demos.kazanas-1@nasa.gov}
\emailAdd{rmohapat@umd.edu}
\emailAdd{vigdor.l.teplitz@nasa.gov}
\emailAdd{yongchao.zhang@ulb.ac.be}

\date{\today}

\abstract{
  We discuss a simple non-supersymmetric model  based on the electroweak  gauge group $SU(2)_L\times SU(2)^\prime\times U(1)_{B-L}$ where the lightest of the right-handed neutrinos, which are part of the leptonic doublet of $SU(2)^\prime$, play the role of a long-lived unstable dark matter with mass in the multi-PeV range. We use a resonant $s$-channel annihilation to obtain the correct thermal relic density and relax the unitarity bound on dark matter mass. In this model, there exists a 3-body dark matter decay mode producing tau leptons and neutrinos, which could be the source for the PeV cascade events observed in the IceCube experiment. The model can be tested with more precise flavor information of the highest-energy neutrino events in future data.
}

\maketitle

\section{Introduction}
The IceCube neutrino telescope has reported 54 ultra-high energy (UHE) neutrino events with deposited energies ranging all the way from 20 TeV to 2 PeV in its 4-year dataset~\cite{Aartsen:2013jdh, Aartsen:2014gkd, Aartsen:2015zva}. This constitutes a $6.4\sigma$ excess over the expected background of atmospheric muons and neutrinos.  The events  seem to be isotropically distributed in the sky, with no statistically significant evidence of point-like sources~\cite{Adrian-Martinez:2015ver, Aartsen:2016tpb}. The specific
origin, spectral shape and flavor composition of these events are currently unknown, but understanding all the observed features so far using conventional astrophysics alone appears to be a challenge~\cite{Anchordoqui:2013dnh, Murase:2014tsa, Kistler:2015ywn}. This has inspired many speculations regarding possible new physics beyond the standard model (BSM) of particle physics that could (partly) explain the origin of these events~\cite{Anchordoqui:2013dnh, Tomar:2015fha}, although the current statistics does not necessarily call for a BSM interpretation yet~\cite{Laha:2013lka, Chen:2013dza, Vincent:2016nut, Barger:2012mz}. Nevertheless, if some peculiar features in the IceCube data, namely, an apparent energy gap just below PeV and a slight excess of events above PeV over the predictions from a single, unbroken power-law astrophysical neutrino flux, as well as the lack of any events near the Glashow resonance of 6.3 PeV~\cite{Glashow:1960zz} and no statistically significant correlation of the neutrino arrival directions with the galactic disk, persist with more data, it might lend support to a simple BSM interpretation in terms of a long lived supermassive (multi-PeV) particle dark matter (DM) decaying into neutrinos~\cite{Feldstein:2013kka, Esmaili:2013gha, Bai:2013nga, Ema:2013nda, Bhattacharya:2014vwa, Higaki:2014dwa, Ema:2014ufa, Rott:2014kfa, Esmaili:2014rma, Fong:2014bsa, Dudas:2014bca, Murase:2015gea, Boucenna:2015tra, Bhattacharya:2014yha, Kopp:2015bfa, Chianese:2016opp, Ko:2015nma}, which is assumed to be the case in this paper. With the conventional direct and indirect detection, as well as collider searches for weakly interacting massive particle (WIMP) DM candidates in the ${\cal O}$(GeV-TeV) range being unsuccessful so far,  it is worthwhile to consider the observational prospects of non-WIMP scenarios, such as the PeV-scale decaying DM alluded to above, for which IceCube offers a unique opportunity. Another general motivation for the presence of a decaying DM component, though not necessarily applicable to our model, is due to the fact that it can possibly alleviate the tension between Planck data and low redshift astronomical measurements~\cite{Berezhiani:2015yta, Anchordoqui:2015lqa}, although the extent of improvement is quite modest~\cite{Poulin:2016nat}, requiring lifetimes comparable with or smaller than the age of the universe and possibly involving only a fraction of the DM. A PeV-scale RH neutrino DM can also be accommodated in leptogenesis models for explaining the matter-antimatter asymmetry~\cite{Anisimov:2008gg, Ahn:2016hhq}.

Since this phenomenon involves neutrinos, it is plausible to surmise that it is related to neutrino mass physics. We take this approach here and present a ultraviolet (UV) complete model, where a heavy right-handed (RH) neutrino, which could be at the root of nonzero neutrino masses, is cosmologically stable enough to play the role of DM in our Universe and, being unstable, produces the energetic neutrinos when it decays.  This model is however very different from the usual type-I seesaw models~\cite{type1a,type1b,type1c,type1d,type1e} for neutrino masses, where the heavy RH neutrinos decay rapidly to the SM Higgs boson and known leptons through their coupling with the light neutrinos, and therefore, cannot qualify as the DM of the Universe, unless the coupling constant is extremely small $\sim {\cal O}(10^{-30})$~\cite{Higaki:2014dwa, Ko:2015nma, Esmaili:2014rma} in which case the sub-eV light neutrino masses cannot be generated from type-I seesaw in the first place. Instead we contemplate that the RH neutrino $N$ has its own $SU(2)^\prime$ gauge interactions (from coupling to the corresponding gauge fields $W^\prime, Z^\prime$) and remains secluded from the SM sector by a softly broken discrete $Z_2$  symmetry so that its decays to lepton plus Higgs are forbidden. It couples only superweakly to the SM sector via tiny scalar, fermion and gauge ($W - W'$ and $Z - Z'$) mixings. These mixings, whose smallness is guaranteed to be natural due to presence of the softly broken $Z_2$ symmetry as well as the large hierarchy between the electroweak (EW) scale and the PeV scale, provide a bridge between the secluded heavy fermion sector and the SM sector. They also explain the ultra-long lifetime for the DM $N$. This is the main new  result of this paper. We then study some implications of this particular UV complete model for the IceCube neutrinos and also discuss the constraints from diffuse gamma ray emission.

This paper is organized as follows: in Section~\ref{sec:model} we provide an outline of the model. In Section~\ref{sec:density} we show how the relic density of the DM $N$ arises in this model.  Section~\ref{sec:neutrino} discusses its various decay modes and how its long lifetime arises. In Section~\ref{sec:ic}, we illustrate how it explains the observed rate for the multi-TeV to PeV neutrinos at IceCube. In Section~\ref{sec:gammaray} we comment on the diffuse gamma ray spectrum associated with this decay. We then conclude with a summary of the results in Section~\ref{sec:conclusion}.

\section{The Model}
\label{sec:model}
The model is based on the gauge group $SU(3)_c \times SU(2)_L \times SU(2)^\prime\times U(1)_{B-L}\times Z_2$, where in addition to the SM fermions, which are singlets under $SU(2)'$ and even under $Z_2$, i.e.
\begin{eqnarray}
{\rm SM~doublets}:&& Q_{L} \ \equiv \ \left(\begin{array}{c} u\\ d\end{array}\right)_{L}: \bigg({\bf 3}, {\bf 2}, {\bf 1}, \frac{1}{3},+\bigg), \nonumber \\
&&\psi_{L} \ \equiv \ \left(\begin{array}{c} \nu \\ e \end{array}\right)_{L}: ({\bf 1}, {\bf 2}, {\bf 1}, -1,+) , \nonumber \\
{\rm SM~ isosinglets}:&& u_R: \bigg({\bf 3}, {\bf 1}, {\bf 1}, \frac{4}{3},+\bigg),~~ d_R: \bigg({\bf 3}, {\bf 1}, {\bf 1}, -\frac{2}{3},+\bigg),~~ \nonumber \\
&&e_R: ({\bf 1}, {\bf 1}, {\bf 1}, -2,+)\, ,
\end{eqnarray}
(here we have neglected the generation indices for simplicity), there are new heavy fermions %(denoted by  $U_{L,R}, D_{L,R}, E_{L,R}$)
which are $SU(2)_L$ singlets with non-zero $U(1)_{B-L}$ charges and transform under the new $SU(2)'$ as follows:
\begin{eqnarray}
{\rm Heavy}~SU(2)^\prime~ {\rm doublets}:&& \mathcal{Q}_{R} \ \equiv \ \left(\begin{array}{c} \mathcal{U} \\ \mathcal{D} \end{array}\right)_{R}: \bigg({\bf 3}, {\bf 1}, {\bf 2}, \frac{1}{3},-\bigg), \nonumber \\ && \Psi_{R} \ \equiv \ \left(\begin{array}{c} N \\ \mathcal{E} \end{array}\right)_{R}: ({\bf 1}, {\bf 1}, {\bf 2}, -1,-), \nonumber \\
{\rm Heavy} ~SU(2)^\prime~{\rm singlets}:&& \mathcal{U}_L:\bigg({\bf 3}, {\bf 1}, {\bf 1}, \frac{4}{3},-\bigg),~~ \mathcal{D}_L:\bigg({\bf 3}, {\bf 1}, {\bf 1}, -\frac{2}{3},-\bigg),~~ \nonumber \\
&& \mathcal{E}_L:({\bf 1}, {\bf 1}, {\bf 1}, -2,-) \,.
\end{eqnarray}
%The $(\mathcal{U}_R,\, \mathcal{D}_R)$ fermions form an $SU(2)^\prime$ doublet $\mathcal{Q}_R$ as do the leptonic fields $(\mathcal{N}_R,\, \mathcal{E}_R)\equiv \Psi_R$, while all the left-handed heavy fermions $\mathcal{U}_L$, $\mathcal{D}_L$ and $\mathcal{E}_L$ are singlets.
We assume that under the $Z_2$ symmetry the heavy multiplets $\mathcal{Q}_R,\, \Psi_R, \mathcal{U}_L,\, \mathcal{D}_L,\, \mathcal{E}_L$ are odd and the SM fermions are even. As a result, the SM and heavy fermions do not form masses of type $\overline{\mathcal{U}}_L u_R$ in the $Z_2$ symmetry limit~\cite{bem1,bem2,bem3,bem4},
%Even though the gauge symmetry and matter fields in our model are to some extent similar to the left-right symmetric models with heavy vector-like fermions~\cite{vectorlike}, the difference is the absence of mass terms of type \blue{$\bar{u}_R \mathcal{U}_L$} connecting the two sectors,
which is crucial to accommodate a long-lived RH neutrino DM candidate in our model. On the other hand, both sectors share common $SU(3)_c$ and $U(1)_{B-L}$ symmetries, with obvious charge assignments, and the electric charge formula is given as in the left-right (LR) symmetric models  by~\cite{Marshak:1979fm, Davidson:1978pm}
\begin{eqnarray}
Q=I_{3L}+I'_{3}+\frac{1}{2} (B-L) \,.
\end{eqnarray}
In this sense, our model duplicates the SM fields except for the common $SU(3)_c$ and $U(1)$ gauge interactions and at this stage similar to the model in Refs.~\cite{bem1,bem2,bem3,bem4}.

The minimal Higgs sector of the model consists of $SU(2)$ and $SU(2)^\prime$ doublets, respectively denoted as
\begin{eqnarray}
&& \chi_\ell \ \equiv \ \left( \begin{matrix} \chi_\ell^{+} \\ \chi_\ell^0 \end{matrix} \right)  : ({\bf 1}, {\bf 2}, {\bf 1}, 1,+) \,, \qquad
\chi'_\ell \ \equiv \  \left( \begin{matrix} \chi_\ell^{\prime \, +} \\ \chi_\ell^{\prime \, 0} \end{matrix} \right):
({\bf 1}, {\bf 1}, {\bf 2}, 1,+) \,, \nonumber \\
&& \chi_q \ \equiv \  \left( \begin{matrix} \chi_q^{+} \\ \chi_q^0 \end{matrix} \right)  :
({\bf 1}, {\bf 2}, {\bf 1}, 1,+) \,, \qquad
\chi'_q \ \equiv \  \left( \begin{matrix} \chi_q^{\prime \, +} \\ \chi_q^{\prime \, 0} \end{matrix} \right):
({\bf 1}, {\bf 1}, {\bf 2}, 1,+) \,,
\end{eqnarray}
which are both lepton-specific in the SM and heavy sector, analogous to the lepton-specific two Higgs doublet model (2HDM)~\cite{Barnett:1983mm, Grossman:1994jb, Goh:2009wg, Branco:2011iw}. The Higgs doublets are chosen to be even under the $Z_2$ symmetry. In the symmetry limit, the Yukawa couplings are given  by the Lagrangian
%\subsection{Completion of LR seesaw model at the TeV scale}
%One ``\blue{hidden}''\footnote{This singlet are introduced mainly to generate the SM fermion masses in a manner of seesaw mechanism and decays dominately via the heavy fermion loops thus it is called ``hidden''. See the sections below.} pure singlet $S$ at TeV scale with even parity can be added to the LR seesaw model to produce the masses of vector like heavy quarks (and leptons) which is the heavy block for generation of the small fermion masses in the SM via the seesaw-like mechanism. Then the complete Yukawa interactions responsible for fermion masses in this model are given by
\begin{eqnarray}
\label{eq:Lyukawa}
- \mathcal{L}_Y & \  \supset \ &
 y_{u}  \bar{Q}_L \tilde{\chi}_{q} u_R
+ y_{d}\bar{Q}_L  \chi_{q} d_R
+ y_{\ell} \bar{\psi}_L  \chi_{\ell} e_R
%+ f \overline{ \psi_R^C} i\sigma_2 \Delta \psi_R %f LL\Delta
\nonumber \\
&& + y'_{u}  \bar{\cal Q}_R \tilde{\chi}'_{q} {\cal U}_L
+ y'_{d}\bar{\cal Q}_R  \chi'_{q} {\cal D}_L
+ y'_{\ell} \bar{\Psi}_R  \chi'_{\ell} {\cal E}_L
%+ (y\to y^\prime,\, f_{L,R}\to \mathcal{F}_{R,L},\, \chi_{\ell,q} \to \chi^\prime_{\ell,q}  )
+ {\rm H.c.} \,,
\end{eqnarray}
where $\tilde{\chi}_q = i\sigma_2\chi_q^*$  and similarly for $\chi_q^\prime$, with $\sigma_2$ being the second Pauli matrix. We give different vacuum expectation values (VEVs) to the doublets $\chi_a$ and $\chi^\prime_a$, i.e.
\begin{eqnarray}
\langle \chi_{\ell,q}^0 \rangle \ = \ v_{\ell,q} \,, \qquad
\langle \chi_{\ell,q}^{\prime\, 0} \rangle \ = \ v^\prime_{\ell,q} \, ,
\end{eqnarray}
with $\sqrt{v_\ell^2 + v_q^2 } \equiv v_{\rm EW} \simeq 174$ GeV and $\sqrt{v'^2_\ell+v'^2_q} \equiv v' \sim {\cal O}$(10 PeV) to accommodate a superheavy DM for explaining the IceCube neutrino events. The VEVs $v_{\ell,q}$ are responsible for the SM charged fermion masses as usual, whereas $v'_{\ell,q}$ make the new charged fermions superheavy. When the ratio of the VEVs  $\tan\beta = \langle \chi_q^0 \rangle / \langle \chi_\ell^0 \rangle \gg 1$, the charged scalar $\chi_\ell^\pm$ decays predominantly into SM leptons. The reason is as follows: as in the most general lepton-specific 2HDMs, the physical charged Higgs boson is a linear combination of the doublets of form $(v_q \chi^{\pm}_\ell-v_\ell\chi^{\pm}_q)$. When the leptonic VEV $v_\ell$ is much smaller than $v_q$, the leptonic Yukawa couplings are enhanced by $y_\ell v_q/y_q v_\ell$. This is crucial to accommodate a leptophilic DM candidate in order to fit the IceCube data~\cite{Boucenna:2015tra}.
%Both comibied lead to the charged Higgs couplings in the leptonic sector to be $\sim y^\prime_\ell v_q$, which is much larger than the corresponding quark coupling which is given by $\sim y^\prime_q v_\ell$.

%To avoid confusion of notation, we emphasize that $U_R, D_R, E_R$ are the right handed partners of the familiar quarks and $u_R, d_R, e_R$ are the partners of the PeV mass heavy quarks and in the symmetry limit, both sectors are distinct and do not mix with one another.

%Even though the gauge symmetry and matter fields in our model are to some extent similar to the left-right symmetric models with heavy vector-like fermions~\cite{vectorlike}, the difference is the absence of mass terms of type \blue{$\bar{u}_R \mathcal{U}_L$} connecting the two sectors,
It is important to point out that this model is different from the conventional LR models with vector-like fermions~\cite{LR1,LR2,LR3,LR4,LR5,LR6,LR7} where the parity symmetry (also a discrete $Z_2$ symmetry) makes the Yukawa couplings equal in the left- and right-handed sectors. In our model instead, we have a different $Z_2$ symmetry which restricts the nature of Yukawa couplings  but not the flavor structure of these couplings. To point out other differences between the two $Z_2$'s,
\begin{itemize}
  \item %Note that parity $Z_2$ will allow the couplings like $\bar{U}_LU_R$ etc among the vectorlike fermions whereas our $Z_2$ does not allow such couplings;
      As mentioned above, the parity $Z_2$ symmetry allow couplings among the SM fermions and the heavy vector-like fermions, whereas in our case, in the symmetry limit, both sectors are distinct and do not mix with one another.
  \item In the parity $Z_2$ case, the SM fermion masses are given by a seesaw formula which depends quadratically on Yukawa couplings in the Lagrangian, whereas in our case, the Yukawa coupling hierarchy depends largely on the VEV ratio $\tan\beta$ as in general lepton-specific 2HDMs~\cite{Barnett:1983mm, Grossman:1994jb, Goh:2009wg, Branco:2011iw}.
\end{itemize}
%For simplicity, we choose the $y^\prime$ couplings to be of order one.

To understand the neutrino masses in the model,  we add two $SU(2)$ triplets
\begin{eqnarray}
\Delta & \ \equiv \ & \left(\begin{array}{cc}
\Delta^+ /\sqrt{2} & \Delta^{++} \\
\Delta^0 & -\Delta^+/\sqrt{2}
\end{array}\right) : ({\bf 1}, {\bf 3}, {\bf 1}, 2,+) \,, \nonumber \\
\Delta' & \ \equiv \ & \left(\begin{array}{cc}
\Delta^{\prime +}/\sqrt{2} & \Delta^{\prime ++} \\
\Delta^{\prime 0} & -\Delta^{\prime +}/\sqrt{2}
\end{array}\right) : ({\bf 1}, {\bf 1}, {\bf 3}, 2,+) \,,
\end{eqnarray}
with a Higgs potential of the form:
\begin{eqnarray}
\label{eqn:potential}
V(\chi_a, \chi_a^\prime, \Delta, \Delta^\prime) & \ \supset \ &
- \mu^2_{ab}\chi^\dagger_a \chi_b
-{\mu^\prime}^2_{ab}\chi^{\prime \dagger}_a\chi^\prime_b
+ M^2 {\rm Tr} ( \Delta^\dagger \Delta )
+ M'^2 {\rm Tr} ( {\Delta^\prime}^\dagger\Delta^\prime ) \nonumber \\
&& + m_{ab} \chi_a^T i\sigma_2 \Delta^\dagger \chi_b
+ m'_{ab} {\chi^\prime_a}^T i\sigma_2 \Delta^{\prime\dagger} \chi^\prime_b \nonumber \\
&& + \lambda_{\chi\chi'} (\chi^T_1\sigma_2\chi_2) (\chi^{\prime\dagger}_1\sigma_2\chi^{\prime *}_2) +  {\rm H.c.} \,,
\end{eqnarray}
with $a=\ell,q$. In the above expression, we have omitted the quartic terms of the form $(\chi^\dagger_a \chi_a)^2$ etc and shown only the terms relevant for heavy and light neutrino masses after spontaneous symmetry breaking by the triplet VEVs
%This potential gives vevs to the neutral members of the scalar multiplets as follows:
\begin{eqnarray}
%<{\chi^\prime}^0>=v^\prime; ~~<{\chi_a}^0>\simeq v_{wk,a}; ~~
\langle \Delta^0 \rangle \ \equiv \ v_L \ \sim \ \frac{m v^2_{\rm EW}}{M^2} \,, \qquad
\langle \Delta^{\prime 0}\rangle \  \equiv \  v^\prime_R \ \sim \ \frac{m' v^{\prime 2}}{M'^ 2} \,.
%\sim v' \,.
\end{eqnarray}
%Here $v_{wk}$ denotes the electroweak scale $\sim 246$ GeV.
We choose the parameters of the model such that $v_L\sim $ eV (corresponding to the soft mass parameter $m \sim 1$ GeV), and  $v'_R \sim 10$ PeV (corresponding to the mass parameters $m',\: M,\: M' \sim 10$ PeV). Given the Yukawa interactions
\begin{eqnarray}
\label{eq:Lyukawa2}
- \mathcal{L}_Y & \ \supset \ &
f \bar{ \psi}_L^C i\sigma_2 \Delta_L \psi_L %f LL\Delta
+ f' \bar{ \Psi}_R^C i\sigma_2 \Delta' \Psi_R
+ {\rm H.c.} \,,
\end{eqnarray}
the $\Delta^{\prime}$ term gives masses to the RH neutrinos of order of few PeV, whereas $v_L$ gives masses to the left-handed neutrinos via the usual type-II seesaw mechanism~\cite{Mohapatra:1980yp,type2a,type2b,type2c,type2d}.

An important feature of this model is that for $\lambda_{\chi\chi'}=0$, there is no $\chi_a^{\pm} - \chi_a^{\prime \, \pm}$, $\Delta^\pm -\Delta^{\prime \, \pm}$ or $W-W^\prime$ mixing at the tree level (even at 1-loop level for the last two). As a result, if the lightest RH neutrino $N$ has mass lower than the other fermions and bosons of the $SU(2)'$-sector, it will be stable in the $Z_2$-symmetric limit. However, when we add  small soft breaking terms to the model of the form
\begin{eqnarray}
\label{eqn:soft}
{\cal L}_{\rm soft} \ = \ \delta_U \bar{\mathcal U}_L u_R +
\delta_D \bar{\mathcal D}_L d_R +
\delta_{\ell} \bar{\mathcal E}_L e_R~+~ {\rm H.c.} \,,
\end{eqnarray}
a small $W - W^\prime$ mixing can be generated at 1-loop level
%via the Feynman diagram in Fig.1. This will
with a magnitude~\cite{Chang:1986bp,Babu:1988yq}
\begin{eqnarray}
\zeta_{WW'} \ \sim \
\frac{gg^\prime \delta_U \delta_D m_t m_b}
{16\pi^2 M_T M_B M^2_{W^\prime}} \, ,
\end{eqnarray}
with $g$ and $g'$ the SM $SU(2)_L$ and $SU(2)'$ gauge couplings respectively, $m_{t,\,b}$ the masses of SM top and bottom quarks, and $M_{T,\,B}$ the masses of heavy top and bottom partner fermions. Since $\delta_{U,\,D}$ are small soft breaking terms, the induced $W-W^\prime$ mixing is small, e.g. $\zeta_{WW'}\sim 10^{-24}\delta_U\delta_D$ GeV$^{-2}$ for PeV-scale $SU(2)'$-breaking. Similarly $\Delta^\pm - \Delta^{\prime\,\pm}$ mixing is also a loop effect and is expected to be of similar order. The dominant Higgs mixing connecting the heavy sector to the light is the $\chi_a^{\pm} - \chi_a^{\prime \, \pm}$ mixing which arises at tree level when $\lambda_{\chi\chi'}\neq 0$ in Eq.~\eqref{eqn:potential}. This mixing is of order $\zeta_{\chi\chi'} \sim \lambda_{\chi \chi'} v_{\rm EW}/v^\prime \sim 10^{-5}\lambda_{\chi\chi'}$.

In the fermion sector, there are also mixings induced by the $\delta_\ell$ term in Eq.~(\ref{eqn:soft}). This mixing only connects the heavy and light charged leptons and is given by $\zeta_{eE}\sim \delta_\ell/v^\prime$. We will see in the subsequent section that both the scalar and fermion mixings are essential to allow a long lifetime for the lightest heavy neutrino $N$, as would be required for understanding the IceCube PeV neutrinos.
%with the latter serves as a decaying DM for the PeV IceCube neutrinos.
%the PeV dark matter $N$ to decay at the cosmological time scale.
%As we will see, this leads to a long lifetime for the lightest primed particle $N$.

%For example, for $\delta_{u,d}\simeq 0.01$ GeV, the lifetime of $N$ is $\sim 10^{27}$ sec, as would be required for understanding the PeV neutrinos~\cite{kusenko}.
%The role $M_\ell$ term in the above equation is chosen to be of the same order as $v_R$ so that  the charged lepton of both sectors have large mixing.

Since there is a clear separation of scales in our model, with only the SM spectrum (in a 2HDM extension) in the infra-red, and all the rest of the spectrum at or above the PeV scale, the two being linked by portal-like interactions, it is instructive to sketch an effective theory from the low-energy point of view and justify the small number coefficients of the phenomenologically important operators. At the effective theory level, our model has the following features added to the SM: the RH neutrino DM $N$, which is  the lightest of the three RH neutrinos in the UV-complete theory presented above and its interactions with SM fields given by
\begin{eqnarray}
{\cal L}_{\rm eff} \ = \ \frac{1}{\Lambda^3}\bar{e}_RN H^\dagger \bar{L} e_R+\frac{1}{\Lambda}LHLH+{\rm H.c.}
\label{eq:eff}
\end{eqnarray}
where $\Lambda$ is of order of the mass of the heavy Higgs boson.
We have chosen the RH neutrino DM to have only leptophilic interactions to be in agreement with data. Also we note that the dimension-5 neutrino mass operator in Eq.~\eqref{eq:eff} arises from a type-II seesaw with a heavy triplet scalar at the PeV scale which has been integrated out in the low-energy effective Lagrangian.

\section{DM Relic density}
\label{sec:density}

For generic thermal relic DM models, there is a generic  upper  limit on the DM mass from the unitarity limit on the annihilation cross section~\cite{mark}. However, as noted already in~\cite{mark} and explicitly demonstrated in this section, the unitarity bound of ${\cal O}$(100) TeV can be relaxed in the case of a resonant annihilation, where the $NN$ annihilation cross section can have a Breit-Wigner enhancement~\cite{Ibe:2008ye}.

In order to determine the relic density of DM, we note that in the early universe, all the heavy particles were in equilibrium with the light SM sector particles due to the $SU(3)_c$ and $U(1)_{B-L}$ gauge  interactions. As the universe cools, the particles of the heavy sector being heavier than the DM $N$, slowly annihilate away leaving the $N$'s in the primordial plasma. As the temperature falls below $M_N$, the DM density gets Boltzmann-suppressed by $e^{-M_N/T}$. The primary annihilation channels to SM particles proceed via particles that connect the two sectors such as  the neutral gauge bosons $Z$ and $ Z'$, which mix at the tree level, and the scalar portal mediated by the $\lambda ( \chi_a^\dagger \chi_a ) {\rm Tr} \left( \Delta^{\prime\dagger} \Delta' \right)$ interaction term in the scalar potential. It is easy to see that since $Z-Z^\prime$ mixing angle is highly suppressed by the VEV ratio $v_{\rm EW}^2/ v'^2$~\cite{LR5}, its contribution to DM annihilation is very small and the Higgs portal dominates, which we consider below.

For calculating the relic density of superheavy DM in the scalar portal, two interactions are relevant. The first one is the Yukawa interaction in Eq.~(\ref{eq:Lyukawa2}). Given the convention $\Delta' \supset v'_R + \Delta^{\prime 0} / \sqrt2$, we obtain the DM mass $M_N = 2 f' v'_R$ and the interaction of DM $N$ to the mediator scalar $f' \bar{N}^C \Delta^{\prime 0} N /\sqrt2 + {\rm H.c.}$ The second one is the quartic scalar coupling $\lambda ( \chi_a^\dagger \chi_a ) {\rm Tr} \left( \Delta^{\prime\dagger} \Delta' \right)$. In the convention of $\chi_a \supset v_{\rm EW} + h/\sqrt2$, we have the interaction term
\begin{eqnarray}
\frac{1}{\sqrt2} \lambda v'_R \Delta^{\prime 0} \left( hh + \phi^0\phi^0 + \phi^+ \phi^- \right) \; ,
\end{eqnarray}
where we have neglected the $\chi_a - \chi'_a$ mixing which is suppressed by $v_{\rm EW}/ v'$, $h$ is the SM Higgs, and $\phi^\pm$ and $\phi^0$ respectively are the longitudinal mode of the SM $W$ and $Z$ bosons. In our case, at the EW scale we have only the four light states above in the scalar sector and all other components are at the PeV scale.\footnote{We need to do the usual fine-tuning of high-scale theories to keep the SM Higgs boson light. We have the freedom of other quartic couplings in the model, such as that appearing in the $(\chi_a'^\dagger \chi'_a ) {\rm Tr} \left( \Delta^{\prime\dagger} \Delta' \right)$ term, so that the other doublet components are heavier than the DM.} We have also neglected the tiny $W - W'$ mixing which arises at 1-loop level in presence of the soft breaking terms in Eq.~(\ref{eqn:soft}).
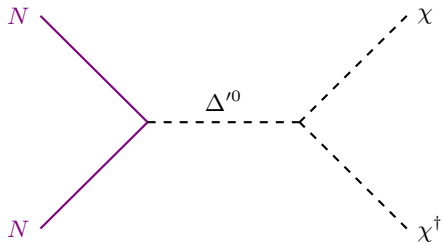
\begin{figure}[t!]
  \centering
  \begin{tikzpicture}[]
\draw[neutrino,thick] (-2.41,1.41)node[left]{{\footnotesize$N$}} -- (-1,0);
\draw[neutrino,thick] (-2.41,-1.41)node[left]{{\footnotesize$N$}} -- (-1,0);
  \draw[dashed,thick](-1,0)--(0,0)node[above]{{\footnotesize$\Delta'^{0}$}}--(1,0);
 \draw[dashed,thick](2.41,1.41)node[right]{{\footnotesize$\chi$}}--(1,0);
 \draw[dashed,thick](2.41,-1.41)node[right]{{\footnotesize$\chi^\dag$}}--(1,0);
  \end{tikzpicture}
  \caption{Feynman diagram responsible for the annihilation of the relic DM $N$ in our model. }
  \label{fig:ann}
\end{figure}

The annihilation processes of interest are (see Figure~\ref{fig:ann})
\begin{eqnarray}
NN \to \Delta^{\prime 0 \, (\ast)} \ \to \ hh,\, \phi^0\phi^0 ,\, \phi^+ \phi^- \,.
\end{eqnarray}
It turns out that the thermally averaged annihilation cross section times velocity is
\begin{eqnarray}
\langle \sigma v \rangle \ = \ 4 \langle \sigma v \rangle_{hh}
\ = \ \frac{ R^2 \lambda^2 }{128 \pi M_N^2} \langle v^2 \rangle \,,
\label{eq:ann1}
\end{eqnarray}
where we have used the DM mass $M_N = 2 f' v'_R$, and
\begin{eqnarray}
R \ = \ \left| \frac{M_N^2}{4M_N^2 - M_{\Delta^{\prime 0}}^2 + iM_{\Delta^{\prime 0}} \Gamma_{\Delta^{\prime 0}}} \right|
\end{eqnarray}
is the resonance enhancement factor. If $R \sim \mathcal{O}(1)$, then there would generally be the unitarity problem with the PeV scale DM, and this cross section is not large enough to reduce the relic density of $N$'s to the desired level. Thus we need a large enhancement factor of $R$ which happens for $M_{\Delta^{\prime 0}} \simeq 2 M_N$. For the sake of simplicity, we assume  all other heavy products such as $W'W'$ are kinematically forbidden and the exact evaluation of $R$ involves only the decay width $\Gamma_{\Delta^{\prime 0}} = \Gamma (\Delta^{\prime 0} \to NN) + 4\Gamma (\Delta^{\prime 0} \to hh)$, where
\begin{eqnarray}
\Gamma (\Delta^{\prime 0} \to NN) & \ = \ & \frac{M_{\Delta^{\prime 0}} M_N^2}{128 \pi v^{\prime 2}_R} \left( 1 - \frac{4M_N^2}{M_{\Delta^{\prime 0}}} \right)^{3/2} \Theta (M_{\Delta^{\prime 0}} - 2M_N) \,, \label{gamma1}\\
\Gamma (\Delta^{\prime 0} \to hh) & \ = \ & \frac{\lambda^2 v^{\prime 2}_R}{16 \pi M_{\Delta^{\prime 0}}} \,.
\end{eqnarray}

%except when there is resonant enhancement of the annihilation cross section with $M_{\Delta^{\prime 0}} \simeq 2M_N$. The enhancement rate for this case is given by:
%assumed the factor $4M_N^2/(4M_N^2 - M_{\Delta^{\prime 0}}^2 + iM_{\Delta^{\prime 0}} \Gamma_{\Delta^{\prime 0}}) \sim \mathcal{O}(1)$, i.e. no fine-tuning in the masses.

%includes the resonance enhancement factor.

\begin{figure}[t!]
  \centering
  \includegraphics[width=8cm]{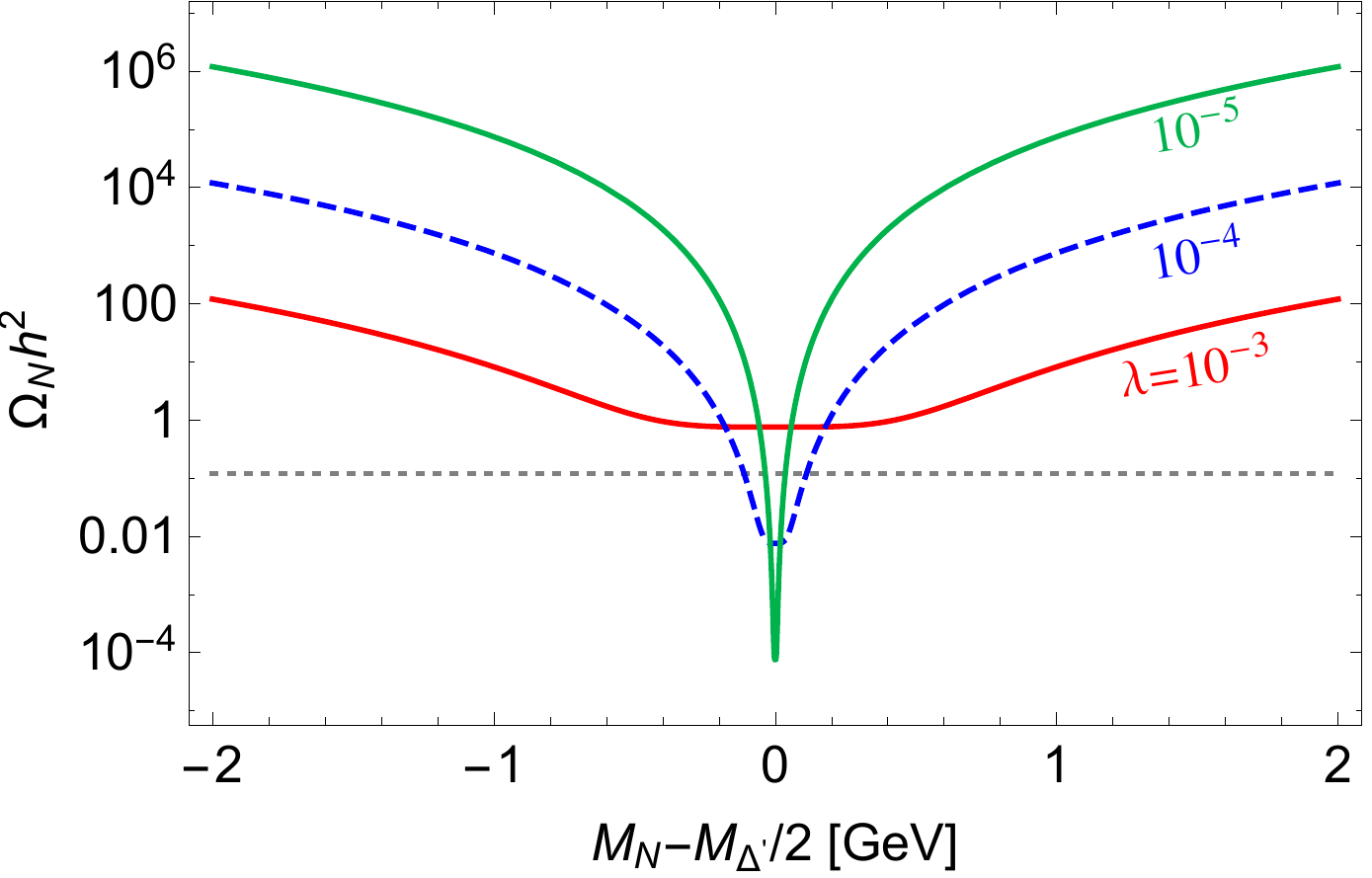}\\
  \caption{Relic density of PeV dark matter $N$ as a function of the resonance parameter $M_N - M_{\Delta^{\prime 0}}/2$ for different values of the quartic coupling $\lambda$. Here we have set $v'_R = M_{\Delta^{\prime 0}} =8$ PeV. The horizontal dashed line shows the observed relic density from Planck data~\cite{Ade:2015xua}.}
  \label{relicdensity}
\end{figure}

%\textbf{Relic density}:
The relic density for the standard thermal relic reads~\cite{Kolb:1990vq}
\begin{eqnarray}
\Omega_N h^2 = \frac{1.07 \times 10^9 \, {\rm GeV}^{-1}}{M_{\rm Pl}} \frac{x_F}{\sqrt{g_\ast}} \frac{1}{a+3b/x_F} \,,
\label{eq:relic1}
\end{eqnarray}
with the Planck mass $M_{\rm Pl} = 1.22 \times 10^{19}$ GeV, $x_F = M_N / T_F \simeq 20$ with $T_F$ being the freeze-out temperature, $g_\ast = 106.75$ the relativistic degrees of freedom at $T_F$, $a$ and $b$ the coefficients in the Taylor expansion $\langle \sigma v \rangle = a + b \langle v^2 \rangle + \mathcal{O} (v^4)$. We consider first a simplified case, where the mediator scalar mass $M_{\Delta^{\prime 0}}$ is very close to but slightly lighter than $2M_N$, then $\Delta^{\prime 0}$ could decay only into the SM Higgs and the longitudinal $W$ and $Z$ components. Assuming $(4M_N^2 - M_{\Delta^{\prime 0}}^2) \ll M_{\Delta^{\prime 0}} \Gamma_{\Delta^{\prime 0}}$, the $p$-wave annihilation cross section in Eq.~\eqref{eq:ann1} becomes
\begin{eqnarray}
\langle \sigma v \rangle \
\simeq \ \frac{\pi M_N^2 }{8\lambda^2 v^{\prime 4}_R} \langle v^2 \rangle
\end{eqnarray}
and the relic density in Eq.~\eqref{eq:relic1} reads
\begin{eqnarray}
\Omega_N h^2 \ \simeq \ \frac{1.07 \times 10^9 \, {\rm GeV}^{-1}}{M_{\rm Pl}} \frac{x_F^2}{\sqrt{g_\ast}} \frac{8\lambda^2 v^{\prime 4}_R}{3\pi M_N^2 } \,.
\label{eq:relic2}
\end{eqnarray}
From Eq.~\eqref{eq:relic2}, it is clear that we can always obtain the right relic density by appropriately choosing the four relevant model parameters $\lambda, \: v'_R,\: M_N$ and $M_{\Delta^{\prime 0}}$. As an illustration, the relic density for a benchmark point in the model parameter space is presented in Figure~\ref{relicdensity} as a function of the deviation from resonance given by $M_N - M_{\Delta^{\prime 0}} /2$. Here we have set $v' = M_{\Delta^{\prime 0}} =8$ PeV and have calculated $\Omega_N h^2$ for different values of the quartic coupling $\lambda$. The horizontal dashed line shows the observed relic density, as measured by Planck~\cite{Ade:2015xua}. As mentioned above, a fine tuning is required for the DM mass $M_N$ (or the mediator mass $M_{\Delta^{\prime 0}}$), i.e. $|M_{\Delta^{\prime 0}} - 2 M_N| < 0.5~{\rm GeV}$, whereas  the quartic coupling $\lambda$ also needs to be small $\lesssim 10^{-3.5}$ in order to reproduce the correct thermal relic abundance. For larger values of $\lambda$, the relic density at the resonance increases, as is evident from Eq.~\eqref{eq:relic2}. For smaller values of $\lambda$, on the other hand, the correct relic density can only be achieved very close to the resonance, which becomes narrower due to the smaller decay width [cf. Eq.~\eqref{gamma1}].

We also point out that in the parameter region away from resonance, where the primordial DM density is higher than desired, the correct value can be obtained by late decay of second lightest RH neutrino to relativistic SM fermions and resulting entropy generation which can cause dilution. We defer discussing the details of the mechanism to a forthcoming paper~\cite{bmz}.

%, e.g.
%\begin{eqnarray}
%\Omega h^2 \ \simeq \ 0.12 \left( \frac{\lambda}{3\times10^{-4}} \right)^2
%\left( \frac{v'_R}{10 \, {\rm PeV}} \right)^4
%\left( \frac{M_N}{5 \, {\rm PeV}} \right)^{-2} \,.
%\end{eqnarray}

%Regarding the exact calculation of the DM relic density, we have only four relevant free parameters, i.e.  the VEV $v'$, the quartic coupling $\lambda$ and the masses $M_{\Delta^{\prime 0}}$ and $M_N$.

\section{DM decay}
\label{sec:neutrino}
%%%%%%%%%%%%%%%%%%%%%%%%%%%%%%
%\textbf{Estimation of the lifetime}:
In this model,  two key pieces of information are important to understand the decay of DM:
\begin{itemize}
  \item All the new particles in the heavy sector are heavier than the RH neutrino DM $N$, which can be achieved by tuning properly the gauge, scalar and Yukawa parameters in the heavy sector.
  \item In the limit of exact $Z_2$ symmetry, interactions between the heavy and light sectors involve two fields from the same sector and therefore in that limit, the $N$'s can only annihilate in pairs but not decay. This is very similar to $R$-parity in supersymmetry~\cite{Farrar:1978xj}.
\end{itemize}
For $N$ to decay to SM fields, we need to invoke soft breaking of $Z_2$ symmetry which can give rise to mixings between $W-W^\prime$, $\Delta^\pm - \Delta^{\prime\,\pm}$ and $\chi_a^\pm - \chi_a^{\prime\,\pm}$. It turns out that the $W-W^\prime$ and $\Delta^\pm - \Delta^{\prime\,\pm}$ mixings are forbidden at the tree level and arise only at loop levels, and are therefore suppressed compared to $\chi_a^\pm - \chi_a^{\prime\,\pm}$ mixing which can arise at the tree level due to the $\lambda_{\chi\chi'}$ term in the scalar potential given by Eq.~(\ref{eqn:potential}). In addition, the small $Z_2$ breaking terms can also induce a $\mathcal{E}_L-e_R$ mixing. %which will allow the decay to proceed.
All these facts then provide a link between the DM $N$ and the SM sector, so that $N$ can decay into light neutrinos, thus giving a potential signal at IceCube. The lifetime of $N$ is resultantly governed by the soft breaking parameters which can be adjusted to be reasonably small to make the lifetime $\tau_N$ of $N$ much longer than the age of the Universe.
%as well as mixing between the heavy and light sector $\mathcal{E}_L-e_R$ fields.

As for the lepton sector, the $Z_2$ conserving and soft breaking terms relevant to the DM decay can be read off from Eqs.~\eqref{eq:Lyukawa} and \eqref{eqn:soft}:
%by explicitly showing the soft $Z_2$ breaking terms.
%The relevant Yukawa couplings for our purpose are:
\begin{eqnarray}
- \mathcal{L}_Y \supset
y_\ell \bar{\psi}_L  \chi_{\ell} e_R +
y'_\ell \bar{\Psi}_R  \chi'_{\ell} \mathcal{E}_L +
\delta_{\ell} \bar{\mathcal E}_L e_R
%h^\prime \bar\Psi_R\chi^\prime E_L + h \bar\psi_L e_R \chi
+ {\rm H.c.}
\end{eqnarray}
After spontaneous symmetry breaking of the $SU(2)'$ and $SU(2)_L$ gauge symmetries at the PeV and EW scales respectively, we obtain the charged lepton mixing $\zeta_{eE}\sim \delta_\ell/v^\prime$.  As for the charged scalar mixing in the doublet sector, i.e. $\chi_a^\pm - \chi_a^{\prime\,\pm}$ mixing,
%we work for simplicity in the limit that the triplets $\Delta_L$ and $\Delta'$ are not involved in.
after symmetry breaking and applying the minimization conditions, we obtain the charged scalar mass terms:
\begin{eqnarray}
\label{eqn:matrix}
\left( \begin{matrix} \chi_{\ell}^+ & \chi_{q}^+ & \chi_\ell^{\prime +} & \chi_q^{\prime +} \end{matrix} \right)
\left( \begin{matrix}
- \lambda_{11} v_R^{\prime \, 2} & \lambda_{12} v_R^{\prime \, 2} & -\lambda_{\chi\chi'} v_2 v'_2 & \lambda_{\chi\chi'} v_2 v'_1 \\
\lambda_{12} v_R^{\prime \, 2} & -\lambda_{22} v_R^{\prime \, 2} & \lambda_{\chi\chi'} v_1 v'_2 & -\lambda_{\chi\chi'} v_1 v'_1 \\
-\lambda_{\chi\chi'} v_2 v'_2 & \lambda_{\chi\chi'} v_1 v'_2 & - \lambda'_{11} v_2^{\prime \, 2} & \lambda'_{12} v_1^{\prime} v_2^\prime \\
\lambda_{\chi\chi'} v_2 v'_1 & -\lambda_{\chi\chi'} v_1 v'_1 & \lambda'_{12} v_1^{\prime} v_2^\prime & - \lambda'_{22} v_1^{\prime \, 2}
\end{matrix} \right)
\left( \begin{matrix} \chi_\ell^- \\ \chi_q^- \\ \chi_\ell^{\prime -} \\ \chi_q^{\prime -} \end{matrix} \right)
\end{eqnarray}
with $\lambda_{ab}, \lambda'_{ab}$ combinations of quartic parameters, VEVs and the soft mass terms $m$ and $m'$ in Eq.~(\ref{eqn:potential}). %These doublets couple also to the triplets $\Delta,\Delta'$, which are not explicitly shown in this mass matrix.
Note that the entries in the left upper $2\times2$ block of Eq.~\eqref{eqn:matrix} are from coupling of $\chi_{\ell,q}$ to the triplet $\Delta'$. The coupling $\lambda_{\chi\chi'}$ bridges the SM and heavy sectors, which is essential for DM $N$ decaying into the SM particles. As expected, two of the charged states in Eq.~\eqref{eqn:matrix} are massless, corresponding to the longitudinal components of the $W$ and $W^{\prime}$ bosons. Gauge invariance demands that none of the two heavy doublets $\chi'_a$ contribute to the SM $W$ mode.
%preventing the otherwise two-body decays into the SM $W$ boson (plus the SM charged lepton via the lepton mixing $\zeta^{Ee}$).
The two heavy states in Eq.~(\ref{eqn:matrix}) are expected to be both at the PeV scale, as $v' \sim v'_R \sim 10$ PeV.

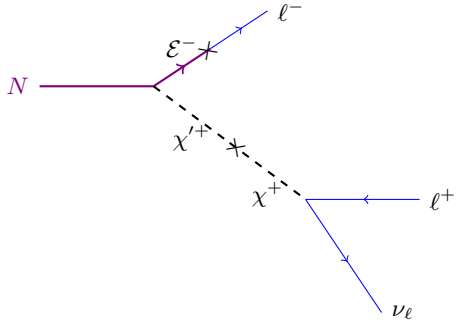
\begin{figure}[t!]
  \def\topdiff{0.25}
  \def\toppos{-1.5}
  \def\vertexstart{-4}
  \def\vertex{\vertexstart+1.5}
  \def\topoffset{0.75}

  \centering
  \begin{tikzpicture}[]
  \draw[neutrino] (\vertexstart-1.5,0)node[left]{{\footnotesize$N$}} -- (\vertexstart,0);
  \draw[electron,thick] (\vertexstart,0)--(\vertexstart+0.375,0.25)node[above]{{\footnotesize$\mathcal{E}^-$}}--(\vertex-0.75,0.5);
  \draw (\vertex-0.55,0.64) node [rotate=35] {$\times$};
  \draw[quark] (\vertex-0.75,0.5)--(\vertex,1)node[right]{{\footnotesize$\ell^-$}};
  \draw[dashed,thick](\vertexstart,0)--(\vertex-1,-0.375)node[below]{{\footnotesize$\chi^{\prime +}$}}--(\vertex-0.5,-0.75);
  \draw (\vertex-0.25,-0.58) node [rotate=55] {$\times$};
  \draw[dashed,thick](\vertex-0.5,-0.75)--(\vertex,-1.125)node[below]{{\footnotesize$\chi^{+}$}}--(\vertex+0.5,-1.5);
  \draw[antiquark](\vertex+0.5,-1.5)--(\vertex+2,-1.5)node[right]{{\footnotesize$\ell^+$}};
  \draw[quark](\vertex+0.5,-1.5)--(\vertex+1.5,-3)node[right]{{\footnotesize$\nu_\ell$}};
  \end{tikzpicture}
  \caption{Feynman diagram responsible for three-body decay of the DM $N$. }
  \label{fig:decay}
\end{figure}

Regarding the DM decay, the two charged scalars $\chi'$ and $\chi$ couple predominantly to the heavy and SM sector respectively and have a mixing of order  $\lambda_{\chi\chi'} v_{\rm EW}/v'$, which gives rise to a 3-body decay of the DM into a light neutrino plus two SM charged leptons: $N\to \ell^- \ell^+ \nu_\ell$, as shown in Figure~\ref{fig:decay}, with the crosses denoting the (heavy-light) scalar and fermion mixings. Actually, one of the prompt charged leptons is produced via its mixing with the heavy charged leptons $\mathcal{E}$ and its flavor depends largely on the texture of $y'_\ell$ and $\delta_\ell$.  For a large $\tan\beta$, the final states are mostly of $\tau$-lepton flavor, as in the case of lepton-specific 2HDM~\cite{Barnett:1983mm, Grossman:1994jb, Goh:2009wg, Branco:2011iw}.

The decay of any DM candidate injects energy into the intergalactic and interstellar medium in the form of quarks, leptons, photons or
neutrinos, which has potential effects on a large number of cosmological and astrophysical observables~\cite{Cline:2013fm, Ibarra:2013cra}. For instance, it can delay recombination and/or contribute to reionization, leading to distortions in the cosmic microwave background (CMB). A recent analysis of the cosmic reionization due to DM decay using the Planck data gives an almost model-independent lower bound on the DM life-time of $\sim 10^{26}$ sec~\cite{Oldengott:2016yjc}, much larger than the actual age of the Universe $\tau_{U}\sim 4\times 10^{17}$ sec. Similar model-independent limits were also obtained using the neutrino flux limits~\cite{PalomaresRuiz:2007ry, Esmaili:2012us}. The soft breaking parameters in 3-body decay of $N$ in our model help to push the DM lifetime to be much longer than the cosmological scale to avoid these constraints.
A rough estimation of the lifetime reads
%\begin{eqnarray}
%\tau_N^{-1} \ = \ \Gamma (N \to e_i e_j \nu) \ \simeq \ \lambda_{\chi\chi'}^2
%\left(\frac{v^4_{\rm EW}\delta^2_\ell}{v^{\prime\, 6}}\right)
%\frac{(y_e y'_e)^4}{1028 \pi^3} \frac{M_N^5}{M_\chi^4} \,.
%\end{eqnarray}
\begin{eqnarray}
\tau_N^{-1} \ & = & \ \Gamma (N \to \ell \ell \nu) \ \simeq \
\frac{\lambda^2_{\chi\chi'}}{192 \pi^3 \tan^2\beta}
\frac{y_\ell^2 }{ y_\ell^{\prime 2}}
\frac{v^4_{\rm EW}\delta^2_\ell}{ v^{\prime 6}}
\frac{ M^5_N}{M^4_\chi} .
\end{eqnarray}
%where the $y'_\ell$ in the denominator is due to the mixing angle  $\
For instance, %$M_N, M_{\chi'}\sim$ PeV, $y_\ell \sim  10^{-2}$, $y'_\ell \sim \mathcal{O}(1)$ and $ \lambda_{\chi\chi'} \sim 10^{-3}$, we get
%\footnote{\blue{Should the final charged leptons be of the $\tau$ flavor, as it has the largest Yukawa coupling of order $10^{-2}$ in the SM???}}
%\footnote{\red{\textbf{One problem with such values of parameters is that: if $y'_e$ is of order $10^{-2}$, then the heavy charged leptons are expected to be lighter than the DM $N$, if the Yukawa coupling $f' \sim 1$ in Eq.(\ref{eq:Lyukawa2}) and the VEVs in the heavy sector $\langle \chi^{\prime 0} \rangle \sim \langle \Delta^{\prime 0} \rangle \sim$ 10 PeV.}}},
%we get\footnote{\blue{YC: I have checked that, the mass matrix for light and heavy charged leptons is
%\begin{eqnarray*}
%\left( \begin{matrix}
  %m_\ell & \delta_\ell \\ \delta_\ell & M_E
%\end{matrix} \right)
%\end{eqnarray*}
%then the heavy-light charged lepton mixing $\zeta_{eE}$ is of order $\delta_\ell / M_E \sim \delta_\ell / v'$. In our case the mediator is a scalar, different from vector-mediated muon decay. However the decay widths and energy spectra are very similar.}}
%\begin{eqnarray}
%\tau_N \  \sim \ (10^{27}\, {\rm sec}) \left(\frac{\rm MeV}{\delta_\ell}\right)^{2} \,  .
%\end{eqnarray}
\begin{eqnarray}
\tau_N
\  &\simeq& \ (10^{27}\, {\rm sec})  \times
\left(\frac{\tan\beta}{60}\right)^{2}
\left(\frac{\lambda_{\chi\chi'}}{3\times10^{-4}}\right)^{-2}
\left(\frac{y_\ell}{ 10^{-2} }\right)^{-2}
\left(\frac{y'_\ell}{ 1 }\right)^{2}
 \nonumber \\
&& \ \times \left(\frac{\delta_\ell}{\rm MeV}\right)^{-2}
\left(\frac{M_N}{ 4 \, {\rm PeV} }\right)^{-5}
\left(\frac{M_\chi}{ 6 \, {\rm PeV} }\right)^{4}
\left(\frac{v'}{ 10 \, {\rm PeV} }\right)^{6} \,.
\end{eqnarray}
Thus, with a small symmetry breaking parameter $\delta_\ell \lesssim $ MeV in Eq.~\eqref{eqn:soft}, we can satisfy the cosmological constraints on our decaying DM scenario. While this is a tiny parameter, since it is a soft breaking of the discrete $Z_2$ symmetry, it is stable under renormalization and is therefore technically natural.

%we have neglected the flavor structure in the final states

%It is worth noting that both the charged scalars $\chi^{\prime \, \pm}$ and $\chi^{\pm}$ from doublets are than the DM $N$, which we can arrange by adjustment of parameters.

%Mediated by the mixing between an off-shell charged scalar $\chi^{\prime {\pm}}$ mixing with $\chi^{\pm}$, the PeV scale RH neutrino $N$ could decay into SM neutrinos plus two charged leptons, i.e.
%\begin{figure}[t!]
%  \centering
%  \includegraphics[width=0.5\textwidth]{Fig1_pevnu.pdf}
%  \caption{Feynman diagram responsible for the decay of the dark matter $N$ }
%  \label{fig:width}
%\end{figure}

\section{Fitting the IceCube Data} \label{sec:ic}

Before getting into details, we give an outline of the argument that shows how this model fits  the observed event rate of the UHE neutrinos on Earth from the decay of $N$, as discussed above. Our arguments are very similar to the phenomenological implications of a generic unstable leptophilic DM~\cite{Boucenna:2015tra}. The DM contribution to energy density in the Universe is roughly 25\% of the critical density $\rho_c=5.5\, {\rm keV\, cm}^{-3}$; so the density of a 4 PeV DM is roughly $n_{\rm DM}\sim 10^{-12}{\rm cm}^{-3}$. If we assume that its lifetime is $\tau_{\rm DM}\sim 10^{27}$ sec, then the probability for each DM to decay is ${\tau_{U}}/{\tau_{\rm DM}} \sim 10^{-10}$. Multiplying it by $n_{\rm DM}$, we get the number density of neutrinos from DM decay to be about $n_\nu \sim 10^{-22}{\rm cm}^{-3}$. To get the flux of neutrinos per steradian on earth, we multiply $n_\nu$ by the velocity of neutrinos $v_\nu \sim c$ (where $c=3\times 10^{10}~{\rm cm\: sec}^{-1}$ is the speed of light) giving
\begin{eqnarray}
E_\nu\Phi_\nu(E_\nu) \ \sim \ \frac{10^6 \, {\rm GeV}}{4\pi \, {\rm sr}} \, n_\nu v_\nu
\ \simeq \ 10^{-7}~{\rm GeV\, cm^{-2}sec^{-1}sr^{-1}} \,,
\end{eqnarray}
which agrees roughly with the flux required to fit the IceCube events at $E\sim 1$ PeV~\cite{Aartsen:2015zva}. Note that for an astrophysical $E^{-2}$ flux as predicted by the Fermi shock accelaration mechanism, the required flux is at the edge of the Waxman-Bahcall bound for optically thin sources~\cite{Waxman:1998yy}. So invoking the decaying DM scenario for PeV events mitigates the situation to some extent.

\begin{figure}[t!]
\centering
\includegraphics[width=8cm]{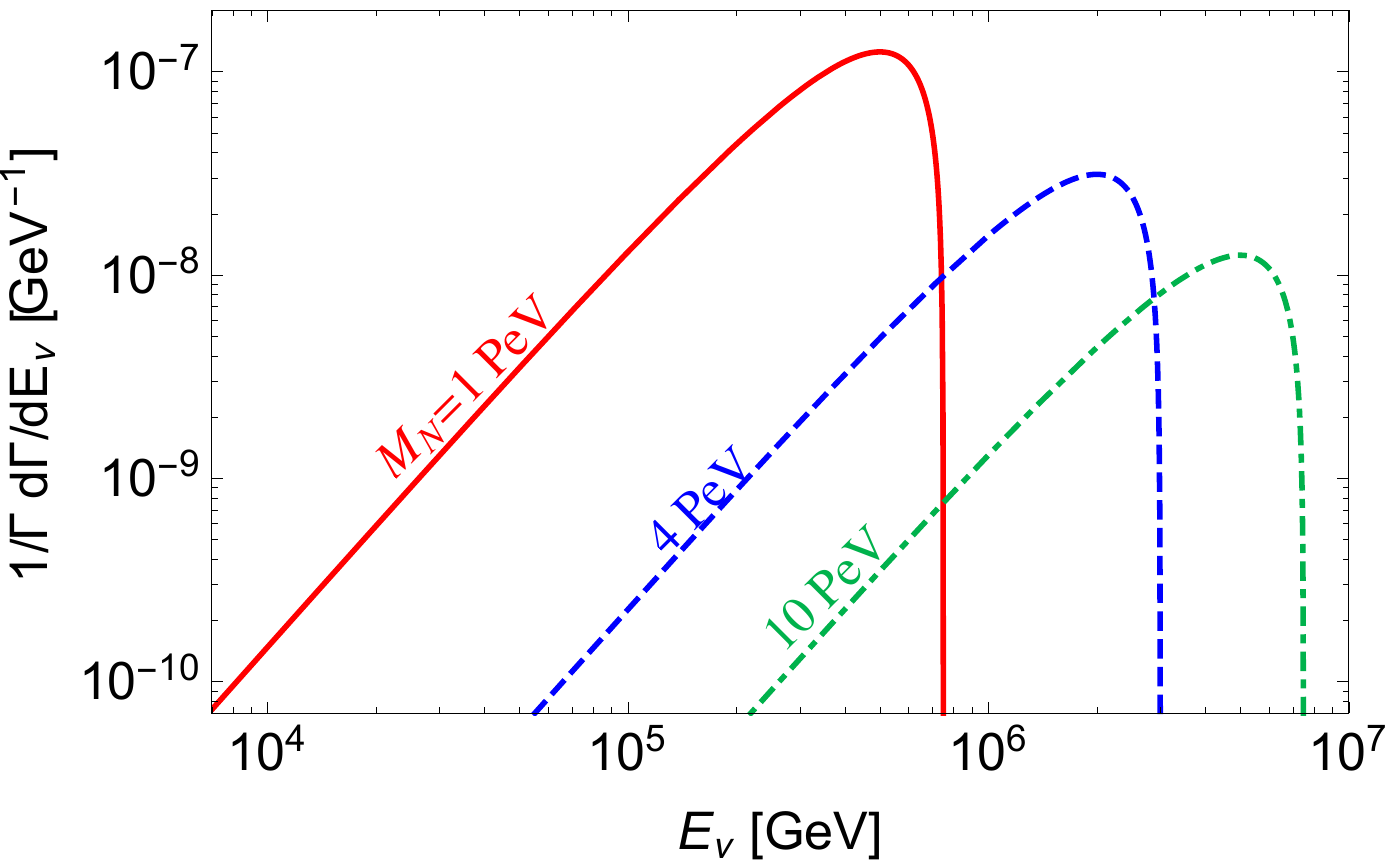}
\caption{The normalized energy distribution of primary neutrinos produced in the 3-body decay  $N\to \tau^+ \tau^- \nu_\tau$ for different values of the DM mass. }
\label{fig:spec}
\end{figure}

In order to do a detailed fitting of the IceCube data, we need the energy distribution of neutrinos in the 3-body decay of the RH neutrino $N\to \tau^+ \tau^- \nu_\tau$. The neutrino energy distribution is similar to the case of electron energy in the muon decay (assuming massless final states) and is given by
%\begin{eqnarray}
%\frac{d\Gamma_N}{dE_\nu} \ = \ G^2_F\zeta^2_{\chi\chi'}\frac{15}{192\pi^3} E^2_\nu (M_N-E_\nu)^2
%\end{eqnarray}
\begin{eqnarray}
\frac{1}{\Gamma_N}\frac{d\Gamma_N}{dE_\nu} \ = \
%\frac{(y_\ell y'_\ell)^2}{128 \pi^3}
%\left(\frac{v^2_{\rm EW}\delta^2_\ell}{v^{\prime\, 4}}\right)
%\frac{M_N^2E_\nu^2}{M_{\chi'}^4}\left(1-\frac{4E_\nu}{3M_N}\right)
\frac{3E_\nu^2}{2M_N^3}\left(1-\frac{4E_\nu}{3M_N}\right) \, .
\end{eqnarray}
%\blue{I find $\left(\frac{v^4_{\rm EW}\delta^2_\ell}{v^{\prime\, 6}}\right)
%\frac{(y_e y'_e)^2}{128 \pi^3} \frac{M_N^2E_\nu^2}{M_{\chi'}^4}\left(1-\frac{4E_\nu}{3M_N}\right)$. Please check.}
In Figure~\ref{fig:spec}, we show the energy spectrum as a function of the true neutrino energy for various representative values of $M_N$. From this, we can infer that the PeV neutrino events at IceCube can be explained by our DM decay scenario with the DM mass of about 4 PeV (see Figure~\ref{fig:event} below).

In practice, the $\tau$-leptons also decay giving rise to secondary neutrinos, as well as electrons, muons and hadrons, which will subsequently lead to more secondary neutrinos in the IceCube detector, thereby raising the tail-end of the spectrum shown in Figure~\ref{fig:spec}. In this respect, the exact neutrino spectrum seen at IceCube will crucially depend on the final-state flavors in the 3-body DM decay, and this feature can in principle be used to probe the flavor structure of the model in future data with more statistics.  It should be noted here that for a larger $W-W'$ mixing at 1-loop level induced by the $\delta_{U,D}$ terms in Eq.~(\ref{eqn:soft}), the 2-body decay modes $N \to W^\pm \ell^\mp$ could also be significant, with the $W$ boson decaying further into SM hadrons or leptons. The large hadronic branching ratio of $W$ gives rise to abundant secondary neutrino emissions from the quarks, which will lead to an almost flat  neutrino flux at energies below PeV~\cite{Esmaili:2013gha}. However, it turns out to be problematic, if a known astrophysical contribution is also added, which by itself provides a good fit to the lower-energy data, and any additional contribution in the lower energy bins seems to be disfavored by the 4-year IceCube data~\cite{Aartsen:2015zva}. Therefore, although in the 2-body DM decay scenario the scalar sector can be made much simpler, we will not consider this case in our analysis and mainly focus on the 3-body decay mentioned above.

To calculate the neutrino flux in our DM decay scenario, we follow the general method outlined in Refs.~\cite{Esmaili:2014rma, Fong:2014bsa, Boucenna:2015tra, Kopp:2015bfa} and consider both galactic (G) and extragalactic (EG) DM components:
\begin{align}
\frac{d\Phi_N(E_\nu)}{dE_\nu} \ = \ \frac{1}{4\pi}\int d\Omega \left( \frac{d\Phi_N^{\rm G}(E_\nu, l, b)}{dE_\nu} +\frac{d\Phi_N^{\rm EG}(E_\nu)}{dE_\nu}  \right) ,
\end{align}
where $\Omega$ is the solid angle and $l,b$ are the longitude and latitude in the galactic coordinate system, respectively. The galactic component can be explicitly written as
\begin{align}
\frac{d\Phi_N^{\rm G}(E_\nu, l, b)}{dE_\nu} \ = \ \frac{1}{4\pi M_N \tau_N} \frac{dN(E_\nu)}{dE_\nu} \int_0^\infty ds\, \rho_N(r(s,l,b)),
\label{gal}
\end{align}
where the distance parameter $s$ is related to the radial distance from the galactic center $r$ by
$r(s,l,b)  =  \sqrt{s^2+R_\odot^2-2sR_\odot \cos b \cos l}$, $R_\odot \simeq 8.5$ kpc is the distance of the Sun from the galactic center, and $\rho_N(r)$ is the DM density profile in Milky Way, for which we assume a Navarro-Frenk-White profile~\cite{Navarro:1996gj}: $\rho_N(r)=\rho_0 (r_0/r)/(1+r/r_0)^2$, with $r_0=20$ kpc and $\rho_0=0.33\, {\rm GeV\, cm}^{-3}$.\footnote{The numerical result for the flux is only weakly dependent on
the DM halo parameters and density profile~\cite{Covi:2009xn}.} The quantity $dN/dE_\nu$ has been computed using the numerical methods outlined in Refs.~\cite{Cirelli:2010xx, Cline:2013fm}, which includes the primary neutrinos and antineutrinos from the DM decay, as well as the secondary ones produced by the $\tau$-lepton decays (including the secondary pion decays).  We have implicitly assumed the sum over all neutrino flavors, whereever applicable.

For the isotropic extragalactic component of the differential flux, we have
\begin{align}
\frac{d\Phi_N^{\rm EG}(E_\nu)}{dE_\nu} \ = \ \frac{\rho_N^{\rm EG}}{4\pi M_N \tau_N} \int_0^\infty \frac{dz}{H(z)}\frac{dN((1+z)E_\nu)}{dE_\nu},
\label{exgal}
\end{align}
where $H(z) = H_0\sqrt{\Omega_\Lambda+\Omega_{\rm m}(1+z)^3}$ is the Hubble rate as a function of the redshift $z$, $H_0=67\,{\rm km\, sec^{-1}{\rm Mpc}^{-1}}$, $\rho_N^{\rm EG} = \Omega_{\rm DM}\rho_c$, and we assume a $\Lambda$CDM cosmology with $\Omega_\Lambda=0.68,\, \Omega_{\rm m}=0.32,\, \Omega_{\rm DM}=0.27$ from the Planck data~\cite{Ade:2015xua}. From Eqs.~\eqref{gal} and \eqref{exgal}, we note that the neutrino flux is inversely proportional to the product of the DM particle mass and
lifetime. Thus for a fixed lifetime, the flux is inversely proportional to the DM
mass due to the lower number density of DM particles.

We also include the standard pion-decay contribution to the flux of astrophysical neutrinos, which could come from known sources like active galactic nuclei~\cite{Murase:2015ndr, Hooper:2016jls} or supernova remnants~\cite{Chakraborty:2015sta}. Since an astrophysical component almost certainly exists and fits the lower-energy part of the IceCube UHE event distribution quite well, it should not be outrightly discarded in favor of an entirely new physics interpretation. As an illustration, we assume a single unbroken power-law astrophysical flux:
\begin{align}
E_\nu^2 \frac{d\Phi_\nu^{\rm astro}(E_\nu)}{dE_\nu} \ = \ \Phi_0 \left(\frac{E_\nu}{100~{\rm TeV}}\right)^{-\gamma},
\end{align}
where $\Phi_0 =2.2\: {\rm GeV\: cm^{-2}\:{\rm sec}^{-1}\:{\rm sr}^{-1}}$ and $\gamma=0.58$ corresponding to the central value of the IceCube best-fit~\cite{Aartsen:2015zva}, assuming $(1:1:1)$ flavor composition on Earth.\footnote{The best-fit solution might be different after we include the decaying DM component~\cite{Boucenna:2015tra}. However, our aim here is not to find the new best-fit solution, since there are several sources of uncertainties for both the astrophysical and DM components. Rather, we use the IceCube best-fit solution just for illustrating the fact that our PeV-scale decaying DM scenario can easily explain the apparent excess of PeV events.} On the other hand, the DM decay in our model produces mostly $\tau$ neutrinos which, after oscillations over astronomical distances, average out to give a flavor ratio of roughly $(4:7:7)$ on Earth~\cite{Learned:1994wg}. The total neutrino flux is given by the sum of the DM decay and astrophysical contributions:
\begin{align}
\frac{d\Phi_\nu(E_\nu)}{dE_\nu} \ = \ \frac{d\Phi_N(E_\nu)}{dE_\nu} + \frac{d\Phi_\nu^{\rm astro}(E_\nu)}{dE_\nu} .
\end{align}

Using this flux and following the analysis method outlined in Refs.~\cite{Chen:2013dza, Vincent:2016nut, Dev:2016uxj}, we compute the number of neutrino events in a given deposited energy bin at IceCube:
\begin{align}
N_{\rm bin} \ = \ T \int_{E^{\rm bin}_{\rm min}}^{E^{\rm bin}_{\rm max}}dE_{\rm dep}(E_\nu)   \frac{d\Phi_\nu(E_\nu)}{dE_\nu} \: A(E_\nu) \, ,
\label{event}
\end{align}
where $T$ is the exposure time, $E_{\rm dep}(E_\nu)$ is the electromagnetic (EM)-equivalent deposited energy for a
given incoming neutrino energy $E_\nu$ in the laboratory frame,  $A$ is the neutrino effective area (for a given flavor), and we have summed over all the neutrino flavors and integrated over the whole sky. Our results are shown in Figure~\ref{fig:event} for a typical benchmark point with $M_N=4$ PeV and $\tau_N=10^{28}$ sec. The background from atmospheric muons and neutrinos, the IceCube data points and the SM best-fit solution (including both charged and neutral current events) are taken from the 4-year IceCube analysis of Ref.~\cite{Aartsen:2015zva}.\footnote{This does not include the latest through-going
track signal with $E_{\rm dep}=2.6\pm 0.3$ PeV~\cite{track}, which most likely would have originated from a $>10$ PeV incoming neutrino. However, with this limited information, we were not able to assess the implications of this event for our decaying DM scenario. For instance, it is essential to know whether this event is accompanied by any shower events in the PeV energy or not. We leave these issues for a future study.} Figure~\ref{fig:event} illustrates the fact that while the low-energy events can be readily explained by an astrophysical component of the neutrino flux, the apparent excess just above PeV energy and the subsequent sharp cut-off can be better understood by invoking a decaying PeV-scale DM hypothesis.

Note that our decaying DM scenario with $\nu_\tau$ final states will mostly produce hadronic showers near the high-energy cut-off, as required to fit the current data. Our fit gives a slight excess in the bin just below PeV, which we believe is still consistent with the IceCube observations, since there are a bunch of lower-energy throughgoing track events, whose true energy could easily be large enough to fill the gap below PeV, given the fact that IceCube can only put a lower limit on the throughgoing muons.

From Figure~\ref{fig:event}, we conclude that although not statistical significant yet, the spectral features in the future IceCube data could be exploited to test the existence of a superheavy DM in our Universe.  In particular, the flavor information could be useful to distinguish our scenario (which predicts mostly $\tau$-flavor final states) from other decaying DM models. A characterisitc feature of the tau-events is the `double-bang' signature; however, for the current IceCube string separation of 120 m,  this feature can be seen only for events with more than 5 PeV energy, which is not kinematically possible for our decaying DM scenario considered in Figure~\ref{fig:event}. Nevertheless, our scenario predicts 17\% more track-events than the SM expectation for a $(1:1:1)$ flavor composition on Earth, and therefore, might be used as a distinguishing feature in future when more data with more accurate flavor information is available. Finally, it should be remarked that the present IceCube is not large enough to test the decay lifetime of $10^{28}$ sec, but there exist other critical multi-messenger tests that are feasible with current and near-future $\gamma$-ray detectors~\cite{Murase:2015gea}.
\begin{figure}[t!]
\centering
\includegraphics[width=12cm]{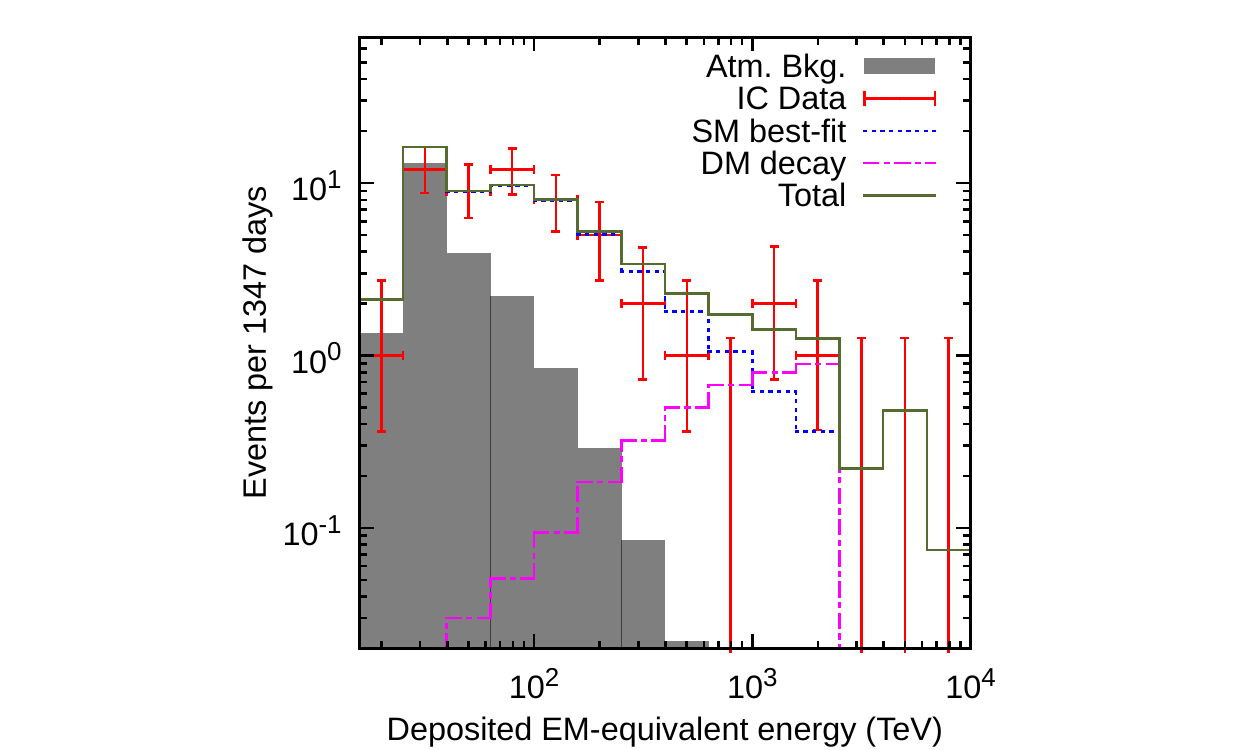}
\caption{The neutrino event distribution for 4-year IceCube with and without the DM decay contribution. For the DM component, we have chosen $M_N=4$ PeV and $\tau_N=10^{28}$ sec.}
\label{fig:event}
\end{figure}

Apart from the excess at PeV-scale, the 4-year IceCube data as shown in Figure~\ref{fig:event} also suggests an apparent excess around 100 TeV. If this becomes statistically more significant with respect to the excess at PeV scale and the apparent energy gap just below the PeV bin disappears, this can in principle be accommodated in our decaying DM scenario by simply choosing a lower value for the dark matter mass in the 300--400 TeV range (along with an appropriate mass spectrum for the other heavy states in the model) and a flatter astrophysical background (e.g. $E^{-2}$). This has already been noted in the literature~\cite{Boucenna:2015tra}, and we do not repeat this analysis here. If more than one excess surface at very different energy scales, it will be difficult to accommodate all of them in a single DM decay scenario, and one might need to invoke a multi-component decaying DM model.

\section{Gamma ray background from DM decay}
\label{sec:gammaray}
%\section{PeV Dark Matter}

One of the features of our model is that the neutrinos will be accompanied by energetic gamma rays from the associated secondary positron production and their subsequent annihilation as they propagate through the intergalactic space, as well as from the final state $\pi^0$'s produced in $\tau$ decay. Naively one would expect that the TeV gamma rays arising in this process will have a flux which is comparable to the neutrino flux i.e.
\begin{eqnarray}
E_\gamma \Phi_\gamma \ \sim \ 10^{-7}\: {\rm GeV}\: {\rm cm}^{-2}\: {\rm sec}^{-1}\: {\rm sr}^{-1} \,.
\end{eqnarray}
Note that this corresponds to an absolute flux of $\Phi_\gamma \sim  10^{-13} \,{\rm cm}^{-2}\,{\rm sec}^{-1}\,{\rm sr}^{-1}$ for a PeV gamma ray. This seems to be just within the current bound on diffuse gamma ray flux at this energy~\cite{turner, Inoue:2012cs,  Murase:2012xs, Gaggero:2015xza, Aartsen:2012gka, Esmaili:2015xpa}. However, one would also expect lower energy gamma rays as the DM decay electrons cool and lose  energy before reaching the Earth. We argue below that those lower energy gamma rays are also well below the current upper limits provided by Fermi-LAT~\cite{Ackermann:2014usa}, HESS~\cite{Abramowski:2014sla}, HAWC~\cite{Abeysekara:2015qba} and  VERITAS~\cite{Archer:2016ein}.

First we note that the electrons will produce gamma rays of energy $E_{\gamma} \simeq m_{\rm DM}/3$ at an
energy flux twice that of neutrinos. One then has to compute the spectrum of resulting photons from the spectrum of injected electrons. If the electrons did not lose energy sufficiently fast compared to their injection time (the age of the Universe) all electrons would be clustered around $E_e \simeq m_{\rm DM}/3$ with their normalization being $N_e \sim Q \times \tau_U$, with $Q$ being the injection rate; then
to compute the photon emission rate we would simply multiply $N_e \, \sigma_T \, c$, where $\sigma_T$ is the Thomson cross section.

However the electrons of PeV energy lose energy fast, mainly on the CMB, barely at the Thomson limit (${\rm PeV} \times E_{\rm CMB} \simeq m_ec^2$). Therefore one has to compute the electron distribution function by dividing the injection rate $Q (\gamma_e)$ by their energy loss time scale. The rate of energy loss of electrons of Lorentz factor $\gamma$ on photons of energy density $\rho_{\rm CMB}$ is given by
\begin{equation}
\dot \gamma_e \ \simeq \ \frac{\rho_{\rm CMB}}{m_ec^2}\, \sigma_{\rm T} c
\gamma_e^2 \ \simeq \ 1.4 \times 10^{-20} \, \gamma_e^2 ~ {\rm sec}^{-1} \,,
\end{equation}
with corresponding time scale $\tau_e \simeq \gamma_e / \dot \gamma_e \simeq 10^{20}/\gamma ~{\rm sec} \simeq  3,000 ~{\rm years}$. So the highest energy electrons lose their energy to photons of similar energy on time scale of 3,000 years. Therefore the amount of PeV gamma rays produced at the source will be the same as the rate of DM decay into neutrinos and electrons. If we fix $\tau_{N} \simeq 10^{28}$ sec to provide neutrino rate in agreement with IceCube data, the photon production rate at PeV energy will be similar.

However, we would like to compute the entire spectrum of the cooling electrons to check if there are discrepancies at other lower energies. The continuous injection of electrons will create differential  power law spectrum of slope $-2$ (because $\dot \gamma \propto \gamma^2$)
\begin{equation}
\frac{dN_e}{d \gamma_e} \ \simeq \ \frac{1}{\dot \gamma_e} \int_{\gamma_e}^{\infty} Q(\gamma^{\prime})\:
d \gamma^{\prime} \ \propto \ \frac{Q}{\gamma_e^2}\:{\rm electrons}\:{\rm sec}^{-1}\:{\rm cm}^{-3}\:{\rm erg}^{-1} \, .
\end{equation}
%The injection rate $Q$ is sharply peaked around $\gamma \simeq m_{DM}/3$. This
%continuous injection and cooling of the electrons produces a power law distribution for
%the electrons of slope -2, i.e. $dN_e/d\gamma \propto \gamma^{-2}$.
%
Inverse Compton scattering on these electrons, upon integration of this distribution with the Compton cross section which has approximately the form $d \sigma/ d \gamma \sim \sigma_{\rm T} \delta(E_{\gamma} - \gamma_e^2 \epsilon)$ (with $\epsilon$ being the soft photon energy that gets up Comptonized), will produce a spectrum of the following form
\begin{eqnarray}
\frac{d\Phi_{\gamma}}{d E_\gamma} = \  \frac{dN_{\gamma_e}}{d E_\gamma}
& \ \propto \ & \frac{dN_e}{d \gamma} \,\sigma_T \, c \,
n_{_{\rm CMB}}(\epsilon) \, \delta(E_{\gamma} - \gamma_e^2 \epsilon) \, d \gamma_e \nonumber \\
& \ \propto \  & E_{\gamma}^{-3/2}\epsilon^{1/2} ~{\rm photons}\:{\rm GeV}^{-1}\: {\rm sec}^{-1}\:
{\rm cm}^{-2}
\end{eqnarray}
and then $E^2_{\gamma} (d\Phi_{\gamma}/d E_{\gamma}) \propto E_{\gamma}^{1/2}\:{\rm GeV\: cm^{-2}\: sec^{-1}}$ with maximum  energy $m_{\rm DM}/3$. The flux at maximum energy will then be twice that of the neutrinos of that energy. However, because the spectrum decreases with  decreasing energy it will not have an impact on the diffuse gamma-ray background (DGRB).

Finally, these gamma rays of the highest energy will be absorbed over distances short compared to the Hubble radius and re-inject electrons to form pairs and a cascade such as that in Refs.~\cite{Kazanas, Zdziarski} . One has to include these in the calculations. The final outcome of such cascades  is to produce a spectrum of photons of slope close to $-2$ (and $E^2_{\gamma}(d\Phi_{\gamma}/d E_{\gamma}) \propto E_{\gamma}^0$) down to a critical photon energy $E_{\gamma,c}$ at which space (at $z\simeq 0$) is transparent to the photon-photon scattering process, i.e. one at which photon-photon opacity $\tau_{\gamma \gamma}(E_{\gamma,\,c}) =1$, and a spectrum $\propto E_{\gamma}^{-3/2}$ (and $E^2_{\gamma}(d \Phi_{\gamma}/dE_{\gamma}) \propto E_{\gamma}^{1/2}$) , still below the level of DGRB, where $E_{\gamma ,c} \simeq 20$ TeV~\cite{Dwek:2012nb}.  Therefore, the observed DGRB at energies $\sim 1 - 50$ GeV is much too bright to be affected by these photons, if the neutrinos from this process are limited not to exceed the observed limits.

Before closing, we also note that a sizable fraction of the photons is produced in the galactic halo, just like for the neutrino flux, and is affected by a more complicated (and partial) absorption/reprocessing, in the ${\cal O}$(100-1000) TeV energy regime. This is typically probed by extensive air shower type of detectors, and could manifest in peculiar anisotropy signals. For a detailed discussion, see e.g. Ref.~\cite{Esmaili:2015xpa}.

\section{Comments and Conclusion}
\label{sec:conclusion}
A few comments are in order regarding the model and its implications:

\begin{itemize}
  \item  If the gauge symmetry $SU(2)'$ is identified as the right-handed gauge group $SU(2)_R$ of the LR symmetric models, then it will be a variation of the conventional LR model with heavy vector-like fermions~\cite{LR1,LR2,LR3,LR4,LR5,LR6,LR7}, with the heavy sector at a higher energy scale. However to have a viable dark matter in this case, we need to increase the scale to much higher values to avoid the 3-body $N$ decay to off-shell $W_R$ mediated by $SU(2)_R$ gauge ineraction and to adjust the RH neutrino Yukawa couplings $f$ appropriately so that the lightest $N$ is lighter than all other fermions of the heavy sector. %It could be made completely parity symmetric prior to symmetry breaking at even superhigh energy scale, e.g. in grand unification theories.

  \item {For the model to work, two fine-tunings are needed: (i) the usual SM Higgs fine-tuning since we have new physics at a PeV scale, and (ii) tuning of the DM mass $2M_N\simeq M_\Delta$ to get the right relic density. The second fine-tuning could be avoided if a non-thermal production mechanism for the right-handed neutrino DM is considered, whereas we need some UV-completion beyond the PeV scale to explain the first one.  If the second right handed neutrino decays later than the lightest one (DM) , then we can avoid this fine tuning.}

  \item We also note that the heavy electron partner of the right-handed DM $N$ can also decay via the Higgs coupling with a life time $\sim 10^{-8}$ sec, so that it would not be present after the Universe's temperature of 10 GeV and hence will not have any impact on the evolution of the Universe around BBN.

  \item In the heavy sector, the heavy quarks will not only form baryonic bound states among themselves but also with the light SM quarks since they share the same QCD. The question then arises if the lightest baryon involving one or more heavy quarks is stable or unstable. In  the limit of exact $Z_2$ symmetry, there will be several stable states i.e. $Qqq, QQq, QQQ$ (where we have not included the proton which in our notation is $qqq$ type). Since the heavy stable baryonic states will have masses in the PeV range, their relic density is likely to far exceed the closure density  unless there are resonant effect like in the case of the heavy DM $N$. One way out of this puzzle would be to introduce soft breaking of the $Z_2$ symmetry by adding mass terms connecting heavy and light quark states e.g. those in Eq.~(\ref{eqn:soft})  and adjust the parameters ($\delta_{U,D}$) to make them decay above the QCD phase transition temperature. For this purpose, a value of $\delta_{U,D}\geq 10^{-4}$ GeV is sufficient.
\end{itemize}

To conclude, we have presented a UV-complete model for PeV-scale decaying dark matter in terms of the lightest right-handed neutrinos which are part of an extra $SU(2)'$ doublet. The cosmological stability of this DM is guaranteed by a discrete $Z_2$ symmetry, which is only softly broken by explicit small mass terms. The apparent excess PeV events at IceCube over the standard expectations from a single unbroken astrophysical power-law flux can be understood as due to the 3-body decay of the PeV-scale DM into tau neutrinos $\nu_\tau$. Our scenario is consistent with other observational constraints, such as the diffuse gamma ray flux and CMB constraints on decaying DM. This model can in principle be tested in future IceCube data with more statistics and more accurate information on the flavor ratio of the observed neutrino events, since it predicts an enhancement of the track events over cascades. The collider and other laboratory tests of this model seem unfeasible, since the non-SM sector lies at or above the PeV scale.

\section*{Acknowledgement}
B.D. is grateful to Pasquale Di Bari and Stefano Morisi for useful discussions on PeV DM at IceCube, and to  the Mainz Institute for Theoretical Physics (MITP) and Munich Institute for Astro- and Particle Physics (MIAPP) for their hospitality and partial support during the completion of this work. Y.Z. would like to thank Julian Heeck for the enlightening discussions, and also Hong-Hao Zhang for his gracious hospitality during the visit at Sun Yat-Sen University where part of the work was done.
The work of B.D. is supported by the DFG grant
RO 2516/5-1. The work of R.N.M. is supported by the US National Science Foundation Grant No. PHY-1315155. Y.Z. would like to thank the IISN and Belgian Science Policy (IAP VII/37) for support. Y.Z. is also grateful to the National Natural Science Foundation of China (NSFC) under Grant No. 11375277 for financial support.


\begin{thebibliography}{99}

\bibitem{Aartsen:2013jdh}
  M.~G.~Aartsen {\it et al.} [IceCube Collaboration],
  %``Evidence for High-Energy Extraterrestrial Neutrinos at the IceCube Detector,''
  Science {\bf 342}, 1242856 (2013)
 % doi:10.1126/science.1242856
  [arXiv:1311.5238 [astro-ph.HE]].
%
\bibitem{Aartsen:2014gkd}
  M.~G.~Aartsen {\it et al.} [IceCube Collaboration],
  %``Observation of High-Energy Astrophysical Neutrinos in Three Years of IceCube Data,''
  Phys.\ Rev.\ Lett.\  {\bf 113}, 101101 (2014)
 % doi:10.1103/PhysRevLett.113.101101
  [arXiv:1405.5303 [astro-ph.HE]].

\bibitem{Aartsen:2015zva}
  M.~G.~Aartsen {\it et al.} [IceCube Collaboration],
  %``The IceCube Neutrino Observatory - Contributions to ICRC 2015 Part II: Atmospheric and Astrophysical Diffuse Neutrino Searches of All Flavors,''
  arXiv:1510.05223 [astro-ph.HE].

\bibitem{Adrian-Martinez:2015ver}
  S.~Adrian-Martinez {\it et al.} [ANTARES and IceCube Collaborations],
  %``First combined search for neutrino point-sources in the Southern Hemisphere with the ANTARES and IceCube neutrino telescopes,''
  arXiv:1511.02149 [hep-ex].

\bibitem{Aartsen:2016tpb}
  M.~G.~Aartsen {\it et al.} [IceCube Collaboration],
  %``Lowering IceCube's Energy Threshold for Point Source Searches in the Southern Sky,''
  arXiv:1605.00163 [astro-ph.HE].



\bibitem{Anchordoqui:2013dnh}
  L.~A.~Anchordoqui {\it et al.},
  %``Cosmic Neutrino Pevatrons: A Brand New Pathway to Astronomy, Astrophysics, and Particle Physics,''
  JHEAp {\bf 1-2}, 1 (2014)
 [arXiv:1312.6587 [astro-ph.HE]].

\bibitem{Murase:2014tsa}
  K.~Murase,
  %``On the Origin of High-Energy Cosmic Neutrinos,''
  AIP Conf.\ Proc.\  {\bf 1666}, 040006 (2015)
%  doi:10.1063/1.4915555
  [arXiv:1410.3680 [hep-ph]].

\bibitem{Kistler:2015ywn}
  M.~D.~Kistler,
  %``Problems and Prospects from a Flood of Extragalactic TeV Neutrinos in IceCube,''
  arXiv:1511.01530 [astro-ph.HE].

%\cite{Tomar:2015fha}
\bibitem{Tomar:2015fha}
  G.~Tomar, S.~Mohanty and S.~Pakvasa,
  %``Lorentz Invariance Violation and IceCube Neutrino Events,''
  JHEP {\bf 1511}, 022 (2015)
  %doi:10.1007/JHEP11(2015)022
  [arXiv:1507.03193 [hep-ph]].
  %%CITATION = doi:10.1007/JHEP11(2015)022;%%
  %1 citations counted in INSPIRE as of 19 Jun 2016

\bibitem{Laha:2013lka}
  R.~Laha, J.~F.~Beacom, B.~Dasgupta, S.~Horiuchi and K.~Murase,
  %``Demystifying the PeV Cascades in IceCube: Less (Energy) is More (Events),''
  Phys.\ Rev.\ D {\bf 88}, 043009 (2013)
  [arXiv:1306.2309 [astro-ph.HE]].

\bibitem{Chen:2013dza}
  C.~Y.~Chen, P.~S.~B.~Dev and A.~Soni,
  %``Standard model explanation of the ultrahigh energy neutrino events at IceCube,''
  Phys.\ Rev.\ D {\bf 89}, 033012 (2014)
  [arXiv:1309.1764 [hep-ph]];
%
%\bibitem{Chen:2014gxa}
%  C.~Y.~Chen, P.~S.~B.~Dev and A.~Soni,
  %``Two-component flux explanation for the high energy neutrino events at IceCube,''
  Phys.\ Rev.\ D {\bf 92}, 073001 (2015)
%  doi:10.1103/PhysRevD.92.073001
  [arXiv:1411.5658 [hep-ph]].


\bibitem{Vincent:2016nut}
  A.~C.~Vincent, S.~Palomares-Ruiz and O.~Mena,
  %``Analysis of the 4-year IceCube HESE data,''
  Phys.\ Rev.\ D {\bf 91}, 103008 (2015)
  [arXiv:1502.02649 [astro-ph.HE]]; arXiv:1605.01556 [astro-ph.HE].

%\cite{Barger:2012mz}
\bibitem{Barger:2012mz}
  V.~Barger, J.~Learned and S.~Pakvasa,
  %``IceCube PeV Cascade Events Initiated by Electron-Antineutrinos at Glashow Resonance,''
  Phys.\ Rev.\ D {\bf 87}, no. 3, 037302 (2013)
  %doi:10.1103/PhysRevD.87.037302
  [arXiv:1207.4571 [astro-ph.HE]].
  %%CITATION = doi:10.1103/PhysRevD.87.037302;%%
  %29 citations counted in INSPIRE as of 19 Jun 2016

%\cite{Glashow:1960zz}
\bibitem{Glashow:1960zz}
  S.~L.~Glashow,
  %``Resonant Scattering of Antineutrinos,''
  Phys.\ Rev.\  {\bf 118}, 316 (1960).
  %doi:10.1103/PhysRev.118.316
  %%CITATION = doi:10.1103/PhysRev.118.316;%%
  %188 citations counted in INSPIRE as of 06 Jun 2016

%\cite{Feldstein:2013kka}
\bibitem{Feldstein:2013kka}
  B.~Feldstein, A.~Kusenko, S.~Matsumoto and T.~T.~Yanagida,
  %``Neutrinos at IceCube from Heavy Decaying Dark Matter,''
  Phys.\ Rev.\ D {\bf 88}, no. 1, 015004 (2013)
  %doi:10.1103/PhysRevD.88.015004
  [arXiv:1303.7320 [hep-ph]].
  %%CITATION = doi:10.1103/PhysRevD.88.015004;%%
  %75 citations counted in INSPIRE as of 06 Jun 2016

%\cite{Esmaili:2013gha}
\bibitem{Esmaili:2013gha}
  A.~Esmaili and P.~D.~Serpico,
  %``Are IceCube neutrinos unveiling PeV-scale decaying dark matter?,''
  JCAP {\bf 1311}, 054 (2013)
  %doi:10.1088/1475-7516/2013/11/054
  [arXiv:1308.1105 [hep-ph]].
  %%CITATION = doi:10.1088/1475-7516/2013/11/054;%%
  %82 citations counted in INSPIRE as of 06 Jun 2016

%\cite{Bai:2013nga}
\bibitem{Bai:2013nga}
  Y.~Bai, R.~Lu and J.~Salvado,
  %``Geometric Compatibility of IceCube TeV-PeV Neutrino Excess and its Galactic Dark Matter Origin,''
  JHEP {\bf 1601}, 161 (2016)
  %doi:10.1007/JHEP01(2016)161
  [arXiv:1311.5864 [hep-ph]].
  %%CITATION = doi:10.1007/JHEP01(2016)161;%%
  %43 citations counted in INSPIRE as of 06 Jun 2016

%\cite{Ema:2013nda}
\bibitem{Ema:2013nda}
  Y.~Ema, R.~Jinno and T.~Moroi,
  %``Cosmic-Ray Neutrinos from the Decay of Long-Lived Particle and the Recent IceCube Result,''
  Phys.\ Lett.\ B {\bf 733}, 120 (2014)
  %doi:10.1016/j.physletb.2014.04.021
  [arXiv:1312.3501 [hep-ph]].
  %%CITATION = doi:10.1016/j.physletb.2014.04.021;%%
  %29 citations counted in INSPIRE as of 06 Jun 2016

%\cite{Bhattacharya:2014vwa}
\bibitem{Bhattacharya:2014vwa}
  A.~Bhattacharya, M.~H.~Reno and I.~Sarcevic,
  %``Reconciling neutrino flux from heavy dark matter decay and recent events at IceCube,''
  JHEP {\bf 1406}, 110 (2014)
  %doi:10.1007/JHEP06(2014)110
  [arXiv:1403.1862 [hep-ph]].
  %%CITATION = doi:10.1007/JHEP06(2014)110;%%
  %26 citations counted in INSPIRE as of 06 Jun 2016

%\cite{Higaki:2014dwa}
\bibitem{Higaki:2014dwa}
  T.~Higaki, R.~Kitano and R.~Sato,
  %``Neutrinoful Universe,''
  JHEP {\bf 1407}, 044 (2014)
  %doi:10.1007/JHEP07(2014)044
  [arXiv:1405.0013 [hep-ph]].
  %%CITATION = doi:10.1007/JHEP07(2014)044;%%
  %21 citations counted in INSPIRE as of 06 Jun 2016

\bibitem{Bhattacharya:2014yha}
  A.~Bhattacharya, R.~Gandhi and A.~Gupta,
  %``The Direct Detection of Boosted Dark Matter at High Energies and PeV events at IceCube,''
  JCAP {\bf 1503}, no. 03, 027 (2015)
%  doi:10.1088/1475-7516/2015/03/027
  [arXiv:1407.3280 [hep-ph]].

%\cite{Ema:2014ufa}
\bibitem{Ema:2014ufa}
  Y.~Ema, R.~Jinno and T.~Moroi,
  %``Cosmological Implications of High-Energy Neutrino Emission from the Decay of Long-Lived Particle,''
  JHEP {\bf 1410}, 150 (2014)
  %doi:10.1007/JHEP10(2014)150
  [arXiv:1408.1745 [hep-ph]].
  %%CITATION = doi:10.1007/JHEP10(2014)150;%%
  %22 citations counted in INSPIRE as of 06 Jun 2016

%\cite{Rott:2014kfa}
\bibitem{Rott:2014kfa}
  C.~Rott, K.~Kohri and S.~C.~Park,
  %``Superheavy dark matter and IceCube neutrino signals: Bounds on decaying dark matter,''
  Phys.\ Rev.\ D {\bf 92}, no. 2, 023529 (2015)
  %doi:10.1103/PhysRevD.92.023529
  [arXiv:1408.4575 [hep-ph]].
  %%CITATION = doi:10.1103/PhysRevD.92.023529;%%
  %31 citations counted in INSPIRE as of 06 Jun 2016

%\cite{Esmaili:2014rma}
\bibitem{Esmaili:2014rma}
  A.~Esmaili, S.~K.~Kang and P.~D.~Serpico,
  %``IceCube events and decaying dark matter: hints and constraints,''
  JCAP {\bf 1412}, no. 12, 054 (2014)
  %doi:10.1088/1475-7516/2014/12/054
  [arXiv:1410.5979 [hep-ph]].
  %%CITATION = doi:10.1088/1475-7516/2014/12/054;%%
  %36 citations counted in INSPIRE as of 06 Jun 2016

%\cite{Fong:2014bsa}
\bibitem{Fong:2014bsa}
  C.~S.~Fong, H.~Minakata, B.~Panes and R.~Z.~Funchal,
  %``Possible Interpretations of IceCube High-Energy Neutrino Events,''
  JHEP {\bf 1502}, 189 (2015)
  %doi:10.1007/JHEP02(2015)189
  [arXiv:1411.5318 [hep-ph]].
  %%CITATION = doi:10.1007/JHEP02(2015)189;%%
  %23 citations counted in INSPIRE as of 06 Jun 2016

%\cite{Dudas:2014bca}
\bibitem{Dudas:2014bca}
  E.~Dudas, Y.~Mambrini and K.~A.~Olive,
  %``Monochromatic neutrinos generated by dark matter and the seesaw mechanism,''
  Phys.\ Rev.\ D {\bf 91}, 075001 (2015)
  %doi:10.1103/PhysRevD.91.075001
  [arXiv:1412.3459 [hep-ph]].
  %%CITATION = doi:10.1103/PhysRevD.91.075001;%%
  %10 citations counted in INSPIRE as of 06 Jun 2016

\bibitem{Kopp:2015bfa}
  J.~Kopp, J.~Liu and X.~P.~Wang,
  %``Boosted Dark Matter in IceCube and at the Galactic Center,''
  JHEP {\bf 1504}, 105 (2015)
%  doi:10.1007/JHEP04(2015)105
  [arXiv:1503.02669 [hep-ph]].

%\cite{Murase:2015gea}
\bibitem{Murase:2015gea}
  K.~Murase, R.~Laha, S.~Ando and M.~Ahlers,
  %``Testing the Dark Matter Scenario for PeV Neutrinos Observed in IceCube,''
  Phys.\ Rev.\ Lett.\  {\bf 115}, no. 7, 071301 (2015)
  %doi:10.1103/PhysRevLett.115.071301
  [arXiv:1503.04663 [hep-ph]].
  %%CITATION = doi:10.1103/PhysRevLett.115.071301;%%
  %25 citations counted in INSPIRE as of 06 Jun 2016

%\cite{Boucenna:2015tra}
\bibitem{Boucenna:2015tra}
  S.~M.~Boucenna, M.~Chianese, G.~Mangano, G.~Miele, S.~Morisi, O.~Pisanti and E.~Vitagliano,
  %``Decaying Leptophilic Dark Matter at IceCube,''
  JCAP {\bf 1512}, no. 12, 055 (2015)
  %doi:10.1088/1475-7516/2015/12/055
  [arXiv:1507.01000 [hep-ph]].
  %%CITATION = doi:10.1088/1475-7516/2015/12/055;%%
  %7 citations counted in INSPIRE as of 06 Jun 2016

\bibitem{Ko:2015nma}
  P.~Ko and Y.~Tang,
  %``IceCube Events from Heavy DM decays through the Right-handed Neutrino Portal,''
  Phys.\ Lett.\ B {\bf 751}, 81 (2015)
%  doi:10.1016/j.physletb.2015.10.021
  [arXiv:1508.02500 [hep-ph]].

\bibitem{Chianese:2016opp}
  M.~Chianese, G.~Miele, S.~Morisi and E.~Vitagliano,
  %``Low energy IceCube data and a possible Dark Matter related excess,''
  Phys.\ Lett.\ B {\bf 757}, 251 (2016)
%  doi:10.1016/j.physletb.2016.03.084
  [arXiv:1601.02934 [hep-ph]].

\bibitem{Berezhiani:2015yta}
  Z.~Berezhiani, A.~D.~Dolgov and I.~I.~Tkachev,
  %``Reconciling Planck results with low redshift astronomical measurements,''
  Phys.\ Rev.\ D {\bf 92}, no. 6, 061303 (2015)
%  doi:10.1103/PhysRevD.92.061303
  [arXiv:1505.03644 [astro-ph.CO]].


\bibitem{Anchordoqui:2015lqa}
  L.~A.~Anchordoqui, V.~Barger, H.~Goldberg, X.~Huang, D.~Marfatia, L.~H.~M.~da Silva and T.~J.~Weiler,
  %``IceCube neutrinos, decaying dark matter, and the Hubble constant,''
  Phys.\ Rev.\ D {\bf 92}, no. 6, 061301 (2015)
%  doi:10.1103/PhysRevD.92.061301
  [arXiv:1506.08788 [hep-ph]].

\bibitem{Poulin:2016nat}
  V.~Poulin, P.~D.~Serpico and J.~Lesgourgues,
  %``A fresh look at linear cosmological constraints on a decaying dark matter component,''
  arXiv:1606.02073 [astro-ph.CO].

\bibitem{Anisimov:2008gg}
  A.~Anisimov and P.~Di Bari,
  %``Cold Dark Matter from heavy Right-Handed neutrino mixing,''
  Phys.\ Rev.\ D {\bf 80}, 073017 (2009)
%  doi:10.1103/PhysRevD.80.073017
  [arXiv:0812.5085 [hep-ph]].

\bibitem{Ahn:2016hhq}
  Y.~H.~Ahn, S.~K.~Kang and C.~S.~Kim,
  %``A Model for Pseudo-Dirac Neutrinos: Leptogenesis and Ultra-High Energy Neutrinos,''
  arXiv:1602.05276 [hep-ph].


\bibitem{type1a}
P. Minkowski, Phys. Lett. B {\bf 67}, 421 (1977).

\bibitem{type1b}
R. N. Mohapatra and G. Senjanovi\'{c}, Phys. Rev. Lett. {\bf 44}, 912 (1980).

\bibitem{type1c}
T. Yanagida, Conf. Proc. C {\bf 7902131}, 95 (1979).

\bibitem{type1d}
M. Gell-Mann, P. Ramond and R. Slansky, Conf. Proc.
{\bf C790927}, 315 (1979)  [arXiv:1306.4669 [hep-th]].

\bibitem {type1e}
S.~L.~Glashow,
  %``The Future of Elementary Particle Physics,''
  NATO Sci.\ Ser.\ B {\bf 61}, 687 (1980).

\bibitem{bem1}
  K.~S.~Babu, D.~Eichler and R.~N.~Mohapatra,
  %``Right-handed Neutrino As Weakly Unstable Dark Matter,''
  Phys.\ Lett.\ B {\bf 226}, 347 (1989).
\bibitem{bem2}
  S.~M.~Barr, D.~Chang and G.~Senjanovic,
  %``Strong CP problem and parity,''
  Phys.\ Rev.\ Lett.\  {\bf 67}, 2765 (1991).
\bibitem{bem3}
  P.~Q.~Hung,
  %``A Model of electroweak-scale right-handed neutrino mass,''
  Phys.\ Lett.\ B {\bf 649}, 275 (2007).
\bibitem{bem4}
  R.~T.~D'Agnolo and A.~Hook,
  %``Finding the Strong CP problem at the LHC,''
  arXiv:1507.00336 [hep-ph].




%\cite{Marshak:1979fm}
\bibitem{Marshak:1979fm}
  R.~E.~Marshak and R.~N.~Mohapatra,
  %``Quark - Lepton Symmetry and B-L as the U(1) Generator of the Electroweak Symmetry Group,''
  Phys.\ Lett.\ B {\bf 91}, 222 (1980).
  %doi:10.1016/0370-2693(80)90436-0
  %%CITATION = doi:10.1016/0370-2693(80)90436-0;%%
  %321 citations counted in INSPIRE as of 06 Jun 2016

%\cite{Davidson:1978pm}
\bibitem{Davidson:1978pm}
  A.~Davidson,
  %``$B^-$l as the Fourth Color, Quark - Lepton Correspondence, and Natural Masslessness of Neutrinos Within a Generalized Ws Model,''
  Phys.\ Rev.\ D {\bf 20}, 776 (1979).
  %doi:10.1103/PhysRevD.20.776
  %%CITATION = doi:10.1103/PhysRevD.20.776;%%
  %85 citations counted in INSPIRE as of 06 Jun 2016

\bibitem{Barnett:1983mm}
  R.~M.~Barnett, G.~Senjanovic, L.~Wolfenstein and D.~Wyler,
  %``Implications of a Light Higgs Scalar,''
  Phys.\ Lett.\ B {\bf 136}, 191 (1984);
%
%\bibitem{Barnett:1984zy}
  R.~M.~Barnett, G.~Senjanovic and D.~Wyler,
  %``Tracking Down Higgs Scalars With Enhanced Couplings,''
  Phys.\ Rev.\ D {\bf 30}, 1529 (1984).

\bibitem{Grossman:1994jb}
  Y.~Grossman,
  %``Phenomenology of models with more than two Higgs doublets,''
  Nucl.\ Phys.\ B {\bf 426}, 355 (1994)
 % doi:10.1016/0550-3213(94)90316-6
  [hep-ph/9401311].

\bibitem{Goh:2009wg}
  H.~S.~Goh, L.~J.~Hall and P.~Kumar,
  %``The Leptonic Higgs as a Messenger of Dark Matter,''
  JHEP {\bf 0905}, 097 (2009)
%  doi:10.1088/1126-6708/2009/05/097
  [arXiv:0902.0814 [hep-ph]].

%\cite{Branco:2011iw}
\bibitem{Branco:2011iw}
  G.~C.~Branco, P.~M.~Ferreira, L.~Lavoura, M.~N.~Rebelo, M.~Sher and J.~P.~Silva,
  %``Theory and phenomenology of two-Higgs-doublet models,''
  Phys.\ Rept.\  {\bf 516}, 1 (2012)
  %doi:10.1016/j.physrep.2012.02.002
  [arXiv:1106.0034 [hep-ph]].
  %%CITATION = doi:10.1016/j.physrep.2012.02.002;%%
  %722 citations counted in INSPIRE as of 09 Jun 2016


\bibitem{LR1}
Z. G. Berezhiani, Phys. Lett. B {\bf 129}, 99 (1983).
\bibitem{LR2}
S. Rajpoot, Mod. Phys. Lett. A {\bf 2}, 307 (1987).
\bibitem{LR3}
A. Davidson and K. C. Wali, Phys. Rev. Lett. {\bf 59}, 393 (1987).
\bibitem{LR4}
K.~S. Babu and R.~N. Mohapatra, Phys. Rev. Lett. {\bf 62}, 1079 (1989).
\bibitem{LR5}
K.~S. Babu and R.~N. Mohapatra, Phys. Rev. D {\bf 41}, 1286 (1990).

%\cite{Mohapatra:2014qva}
\bibitem{LR6}
  R.~N.~Mohapatra and Y.~Zhang,
  %``TeV Scale Universal Seesaw, Vacuum Stability and Heavy Higgs,''
  JHEP {\bf 1406}, 072 (2014)
  %doi:10.1007/JHEP06(2014)072
  [arXiv:1401.6701 [hep-ph]].
  %%CITATION = doi:10.1007/JHEP06(2014)072;%%
  %8 citations counted in INSPIRE as of 06 Jun 2016

%\cite{Dev:2015vjd}
\bibitem{LR7}
  P.~S.~B.~Dev, R.~N.~Mohapatra and Y.~Zhang,
  %``Quark Seesaw, Vectorlike Fermions and Diphoton Excess,''
  JHEP {\bf 1602}, 186 (2016)
  %doi:10.1007/JHEP02(2016)186
  [arXiv:1512.08507 [hep-ph]].
  %%CITATION = doi:10.1007/JHEP02(2016)186;%%
  %80 citations counted in INSPIRE as of 06 Jun 2016

\bibitem{Mohapatra:1980yp}
  R.~N.~Mohapatra and G.~Senjanovic,
  %``Neutrino Masses and Mixings in Gauge Models with Spontaneous Parity Violation,''
  Phys.\ Rev.\ D {\bf 23}, 165 (1981).

\bibitem{type2a}
M. Magg and C. Wetterich, Phys. Lett. {\bf B 94}, 61 (1980).

\bibitem{type2b}
J.~Schechter and J.~W.~F.~Valle,
%   %``Neutrino Masses in SU(2) x U(1) Theories,''
  Phys.\ Rev.\ D {\bf 22}, 2227 (1980).

\bibitem{type2c}
T. P. Cheng and L.-F. Li, Phys. Rev. {\bf D 22}, 2860 (1980).

\bibitem{type2d}
G. Lazarides, Q. Shafi and C. Wetterich, Nucl. Phys. {\bf B 181}, 287 (1981).

%\cite{Chang:1986bp}
\bibitem{Chang:1986bp}
  D.~Chang and R.~N.~Mohapatra,
  %``Small and Calculable Dirac Neutrino Mass,''
  Phys.\ Rev.\ Lett.\  {\bf 58}, 1600 (1987).
  %doi:10.1103/PhysRevLett.58.1600
  %%CITATION = doi:10.1103/PhysRevLett.58.1600;%%
  %148 citations counted in INSPIRE as of 06 Jun 2016

%\cite{Babu:1988yq}
\bibitem{Babu:1988yq}
  K.~S.~Babu and X.~G.~He,
  %``Dirac Neutrino Masses As Two Loop Radiative Corrections,''
  Mod.\ Phys.\ Lett.\ A {\bf 4}, 61 (1989).
  %doi:10.1142/S0217732389000095
  %%CITATION = doi:10.1142/S0217732389000095;%%
  %24 citations counted in INSPIRE as of 06 Jun 2016

\bibitem{mark}
  K.~Griest and M.~Kamionkowski,
  %``Unitarity Limits on the Mass and Radius of Dark Matter Particles,''
  Phys.\ Rev.\ Lett.\  {\bf 64}, 615 (1990).

\bibitem{Ibe:2008ye}
  M.~Ibe, H.~Murayama and T.~T.~Yanagida,
  %``Breit-Wigner Enhancement of Dark Matter Annihilation,''
  Phys.\ Rev.\ D {\bf 79}, 095009 (2009)
%  doi:10.1103/PhysRevD.79.095009
  [arXiv:0812.0072 [hep-ph]].

\bibitem{Kolb:1990vq}
  E.~W.~Kolb and M.~S.~Turner,
  %``The Early Universe,''
  Front.\ Phys.\  {\bf 69}, 1 (1990).

\bibitem{Ade:2015xua}
  P.~A.~R.~Ade {\it et al.} [Planck Collaboration],
  %``Planck 2015 results. XIII. Cosmological parameters,''
  arXiv:1502.01589 [astro-ph.CO].

  \bibitem{bmz} P. S. Bhupal Dev, R. N. Mohapatra and Yongchao Zhang, (to appear).

\bibitem{Farrar:1978xj}
  G.~R.~Farrar and P.~Fayet,
  %``Phenomenology of the Production, Decay, and Detection of New Hadronic States Associated with Supersymmetry,''
  Phys.\ Lett.\ B {\bf 76}, 575 (1978).
%  doi:10.1016/0370-2693(78)90858-4

\bibitem{Cline:2013fm}
  J.~M.~Cline and P.~Scott,
  %``Dark Matter CMB Constraints and Likelihoods for Poor Particle Physicists,''
  JCAP {\bf 1303}, 044 (2013)
  Erratum: [JCAP {\bf 1305}, E01 (2013)]
%  doi:10.1088/1475-7516/2013/03/044, 10.1088/1475-7516/2013/05/E01
  [arXiv:1301.5908 [astro-ph.CO]].

\bibitem{Ibarra:2013cra}
  A.~Ibarra, D.~Tran and C.~Weniger,
  %``Indirect Searches for Decaying Dark Matter,''
  Int.\ J.\ Mod.\ Phys.\ A {\bf 28}, 1330040 (2013)
%  doi:10.1142/S0217751X13300408
  [arXiv:1307.6434 [hep-ph]].

\bibitem{Oldengott:2016yjc}
  I.~M.~Oldengott, D.~Boriero and D.~J.~Schwarz,
  %``Reionization and dark matter decay,''
  arXiv:1605.03928 [astro-ph.CO].

\bibitem{PalomaresRuiz:2007ry}
  S.~Palomares-Ruiz,
  %``Model-Independent Bound on the Dark Matter Lifetime,''
  Phys.\ Lett.\ B {\bf 665}, 50 (2008)
 % doi:10.1016/j.physletb.2008.05.040
  [arXiv:0712.1937 [astro-ph]].

\bibitem{Esmaili:2012us}
  A.~Esmaili, A.~Ibarra and O.~L.~G.~Peres,
  %``Probing the stability of superheavy dark matter particles with high-energy neutrinos,''
  JCAP {\bf 1211}, 034 (2012)
%  doi:10.1088/1475-7516/2012/11/034
  [arXiv:1205.5281 [hep-ph]].


\bibitem{Navarro:1996gj}
  J.~F.~Navarro, C.~S.~Frenk and S.~D.~M.~White,
  %``A Universal density profile from hierarchical clustering,''
  Astrophys.\ J.\  {\bf 490}, 493 (1997)
%  doi:10.1086/304888
  [astro-ph/9611107].


\bibitem{Covi:2009xn}
  L.~Covi, M.~Grefe, A.~Ibarra and D.~Tran,
  %``Neutrino Signals from Dark Matter Decay,''
  JCAP {\bf 1004}, 017 (2010)
 % doi:10.1088/1475-7516/2010/04/017
  [arXiv:0912.3521 [hep-ph]].

\bibitem{Waxman:1998yy}
  E.~Waxman and J.~N.~Bahcall,
  %``High-energy neutrinos from astrophysical sources: An Upper bound,''
  Phys.\ Rev.\ D {\bf 59}, 023002 (1999)
%  doi:10.1103/PhysRevD.59.023002
  [hep-ph/9807282].



\bibitem{Cirelli:2010xx}
  M.~Cirelli {\it et al.},
  %``PPPC 4 DM ID: A Poor Particle Physicist Cookbook for Dark Matter Indirect Detection,''
  JCAP {\bf 1103}, 051 (2011)
  Erratum: [JCAP {\bf 1210}, E01 (2012)]
%  doi:10.1088/1475-7516/2012/10/E01, 10.1088/1475-7516/2011/03/051
  [arXiv:1012.4515 [hep-ph]].

\bibitem{Murase:2015ndr}
  K.~Murase,
  %``Active Galactic Nuclei as High-Energy Neutrino Sources,''
  arXiv:1511.01590 [astro-ph.HE].

\bibitem{Hooper:2016jls}
  D.~Hooper,
  %``A Case for Radio Galaxies as the Sources of IceCube's Astrophysical Neutrino Flux,''
  arXiv:1605.06504 [astro-ph.HE].

\bibitem{Chakraborty:2015sta}
  S.~Chakraborty and I.~Izaguirre,
  %``Diffuse neutrinos from extragalactic supernova remnants: Dominating the 100 TeV IceCube flux,''
  Phys.\ Lett.\ B {\bf 745}, 35 (2015)
%  doi:10.1016/j.physletb.2015.04.032
  [arXiv:1501.02615 [hep-ph]].

\bibitem{Learned:1994wg}
  J.~G.~Learned and S.~Pakvasa,
  %``Detecting tau-neutrino oscillations at PeV energies,''
  Astropart.\ Phys.\  {\bf 3}, 267 (1995)
 % doi:10.1016/0927-6505(94)00043-3
  [hep-ph/9405296, hep-ph/9408296].

\bibitem{Dev:2016uxj}
  P.~S.~B.~Dev, D.~K.~Ghosh and W.~Rodejohann,
  %``R-parity Violating Supersymmetry at IceCube,''
  arXiv:1605.09743 [hep-ph].

\bibitem{track} S. Schoenen and L. Raedel [IceCube Collaboration],
Astronomer's
Telegram {\bf 7856}, 1 (2015).



\bibitem{turner}  M.~T.~Ressell and M.~S.~Turner,
  %``The Grand Unified Photon Spectrum: A Coherent View of the Diffuse Extragalactic Background Radiation,''
  Comments Astrophys.\  {\bf 14}, 323 (1990)
  [Bull.\ Am.\ Astron.\ Soc.\  {\bf 22}, 753 (1990)].

\bibitem{Murase:2012xs}
  K.~Murase and J.~F.~Beacom,
  %``Constraining Very Heavy Dark Matter Using Diffuse Backgrounds of Neutrinos and Cascaded Gamma Rays,''
  JCAP {\bf 1210}, 043 (2012)
 % doi:10.1088/1475-7516/2012/10/043
  [arXiv:1206.2595 [hep-ph]].

\bibitem{Inoue:2012cs}
  Y.~Inoue and K.~Ioka,
  %``Upper Limit on the Cosmological Gamma-ray Background,''
  Phys.\ Rev.\ D {\bf 86}, 023003 (2012)
%  doi:10.1103/PhysRevD.86.023003
  [arXiv:1206.2923 [astro-ph.HE]].

\bibitem{Aartsen:2012gka}
  M.~G.~Aartsen {\it et al.} [IceCube Collaboration],
  %``Search for Galactic PeV Gamma Rays with the IceCube Neutrino Observatory,''
  Phys.\ Rev.\ D {\bf 87}, no. 6, 062002 (2013)
%  doi:10.1103/PhysRevD.87.062002
  [arXiv:1210.7992 [astro-ph.HE]].

\bibitem{Gaggero:2015xza}
  D.~Gaggero, D.~Grasso, A.~Marinelli, A.~Urbano and M.~Valli,
  %``The gamma-ray and neutrino sky: A consistent picture of Fermi-LAT, Milagro, and IceCube results,''
  Astrophys.\ J.\  {\bf 815}, no. 2, L25 (2015)
 % doi:10.1088/2041-8205/815/2/L25
  [arXiv:1504.00227 [astro-ph.HE]].

\bibitem{Esmaili:2015xpa}
  A.~Esmaili and P.~D.~Serpico,
  %``Gamma-ray bounds from EAS detectors and heavy decaying dark matter constraints,''
  JCAP {\bf 1510}, no. 10, 014 (2015)
%  doi:10.1088/1475-7516/2015/10/014
  [arXiv:1505.06486 [hep-ph]].



\bibitem{Ackermann:2014usa}
  M.~Ackermann {\it et al.} [Fermi-LAT Collaboration],
  %``The spectrum of isotropic diffuse gamma-ray emission between 100 MeV and 820 GeV,''
  Astrophys.\ J.\  {\bf 799}, 86 (2015)
%  doi:10.1088/0004-637X/799/1/86
  [arXiv:1410.3696 [astro-ph.HE]].

\bibitem{Abramowski:2014sla}
  A.~Abramowski {\it et al.} [HESS Collaboration],
  %``Diffuse Galactic gamma-ray emission with H.E.S.S,''
  Phys.\ Rev.\ D {\bf 90}, no. 12, 122007 (2014)
%  doi:10.1103/PhysRevD.90.122007
  [arXiv:1411.7568 [astro-ph.HE]].

\bibitem{Abeysekara:2015qba}
  A.~U.~Abeysekara {\it et al.} [HAWC Collaboration],
  %``Search for TeV Gamma-Ray Emission from Point-like Sources in the Inner Galactic Plane with a Partial Configuration of the HAWC Observatory,''
  Astrophys.\ J.\  {\bf 817}, no. 1, 3 (2016)
%  doi:10.3847/0004-637X/817/1/3
  [arXiv:1509.05401 [astro-ph.HE]].

\bibitem{Archer:2016ein}
  A.~Archer {\it et al.} [VERITAS Collaboration],
  %``TeV Gamma-ray Observations of The Galactic Center Ridge By VERITAS,''
  Astrophys.\ J.\  {\bf 821}, no. 2, 129 (2016)
%  doi:10.3847/0004-637X/821/2/129
  [arXiv:1602.08522 [astro-ph.HE]].

\bibitem{Kazanas}
D. Kazanas,  Astrophys.\ J.\ {\bf 87}, 112 (1984).

\bibitem{Zdziarski}  A. Zdziarski and A. P. Lightman, Ap. J. {\bf 294}, L79 (1985).

%\cite{Dwek:2012nb}
\bibitem{Dwek:2012nb}
  E.~Dwek and F.~Krennrich,
  %``The Extragalactic Background Light and the Gamma-ray Opacity of the Universe,''
  Astropart.\ Phys.\  {\bf 43}, 112 (2013)
  %doi:10.1016/j.astropartphys.2012.09.003
  [arXiv:1209.4661 [astro-ph.CO]].
  %%CITATION = doi:10.1016/j.astropartphys.2012.09.003;%%
  %72 citations counted in INSPIRE as of 06 Jun 2016


%%%%%%%%%%%%%%%%%%%%%%%%%%%%%%%%%%%%%%%%%%%%%%%%%%%%%%%%%%%%
%%%%%%%%%%%%%%%%%%%%%%%%%%%%%%%%%%%%%%%%%%%%%%%%%%%%%%%%%%%%
%%%%%%%%%%%%%%%%%%%%%%%%%%%%%%%%%%%%%%%%%%%%%%%%%%%%%%%%%%%%





\end{thebibliography}
\end{document}